\documentclass[12pt,a4paper]{article}
\pdfoutput=1
\usepackage{graphicx}
\usepackage[T1]{fontenc}
\usepackage[utf8]{inputenc}
\usepackage{textcomp}
\usepackage[sc,osf]{mathpazo}
\usepackage{a4wide}  
\usepackage{latexsym,amsthm,amsfonts,amsmath,mathrsfs,amssymb}
\usepackage{dsfont}
\usepackage{accents}
\usepackage[nosort]{cite}
\usepackage{booktabs} 
\usepackage[unicode,implicit]{hyperref}
\hypersetup{%
  pdftitle    = {On the supersymmetric solutions of the Heterotic Superstring
    effective action}
  pdfkeywords = {supersymmetry, gravity, supergravity, strings, superstrings,
    solutions, unbroken supersymmetry, Killing spinors, alpha prime
    corrections, bilinears, Fierz identities, bilinear algebra, spin7
    structure, g2 structure, soliton, instanton, BPST, octonionic}
  pdfauthor   = {Andrea Fontanella and Tomas Ortin},
  plainpages  = true,
  colorlinks  = true,
  citecolor   = blue,
  urlcolor    = red,
  linkcolor   = black
}
\newcommand{\hepth}[1]{{\tt
\href{http://www.arXiv.org/abs/hep-th/#1}{hep-th/#1}}}

\newcommand{\arxiv}[1]{{\tt arXiv:\href{http://www.arXiv.org/abs/#1}{#1}}}
\newcommand{\doi}[1]{{\tt DOI:\href{http://dx.doi.org/#1}{#1}}}

\makeatletter
\@addtoreset{equation}{section}
\makeatother

\begin{document}

\begin{flushright}
\small
HU-EP-19/30 \\
IFT-UAM/CSIC-19-117\\
\texttt{arXiv:1910.08496 [hep-th]}\\
May 20\textsuperscript{th}, 2020\\
\normalsize
\end{flushright}

\vspace{1cm}

\begin{center}

  {\Large {\bf On the supersymmetric solutions\\[.5cm]
      of the\\[.5cm]
      Heterotic Superstring effective action}}

\vspace{1.5cm}

\renewcommand{\thefootnote}{\alph{footnote}}

{\sl\large Andrea Fontanella$^{1,2}$}${}^{,}$\footnote{Email: {\tt andrea.fontanella[at]physik.hu-berlin.de}}
{\sl\large and Tom\'{a}s Ort\'{\i}n${}^{2}$}${}^{,}$\footnote{Email: {\tt tomas.ortin[at]csic.es}}

\setcounter{footnote}{0}
\renewcommand{\thefootnote}{\arabic{footnote}}

\vspace{1cm}

\textit{$^{1}$Institut f\"ur Physik, Humboldt-Universit\"at zu Berlin, \\
IRIS Geb\"aude, Zum Grossen Windkanal 6, 12489 Berlin, Germany}

\vspace{5mm}
{\it $^{2}$Instituto de F\'{\i}sica Te\'orica UAM/CSIC\\
C/ Nicol\'as Cabrera, 13--15,  C.U.~Cantoblanco, E-28049 Madrid, Spain}

\vspace{1cm}


{\bf Abstract}
\end{center}
\begin{quotation}
  {\small We consider the effective action of the Heterotic Superstring to
    first order in $\alpha'$ and derive the necessary and sufficient
    conditions that a field configuration has to satisfy in order to admit at
    least one Killing spinor using the spinor bilinear method in an arbitrary
    spinorial basis and corresponding arbitrary gamma matrices.  As a previous
    step in this derivation, we compute the complete spinor bilinear algebra
    using the Fierz identities, obtaining as a by-product the algebra
    satisfied by the Spin(7) structure contained in the bilinears in an
    arbitrary basis. We find the off-shell relations existing between the
    bosonic equations of motion evaluated on supersymmetric field
    configurations using the Killing Spinor Identities instead of the (far
    more complicated) integrability conditions of the Killing Spinor Equations
    as it is common in the literature. We show how to include the
    Kalb-Ramond's Bianchi identity in the Killing Spinor Identities.}
\end{quotation}

\newpage
\pagestyle{plain}

\tableofcontents

\newpage

\section{Introduction}

The construction and study of the classical solutions of a theory always
provides a great deal of information about its properties and predictions. The
fundamental r\^ole played by the Schwarzschild solution in the conceptual
development of General Relativity, as well as in more mundane computations of
testable predictions of this theory, is a very clear example that cannot be
overstated. For these reasons, the construction and study of solutions of the
Superstring Theory effective action (compactification backgrounds, $pp$-wave
backgrounds, black holes, cosmological models) has been a very active area of
research for almost 30 years and it is not surprising that some of the
solutions found, such as the 3-charge black hole
\cite{Tseytlin:1996as,Cvetic:1995bj} used by Strominger and Vafa to compare
the Bekenstein-Hawking entropy with that obtained by the first microstate
counting in Ref.~\cite{Strominger:1996sh}, have also had a huge impact in the
development of Superstring Theory. Near-horizon geometries, $pp$-waves and
other Penrose limits of solutions provide further examples.

The methods used to construct new solutions of Superstring Theory based on
dualities and supersymmetry have probably been the most fruitful ones.
Dualities transform solutions into solutions, sometimes with very different
properties. The original solutions are required to satisfy only a minimal
number of conditions such as the existence of isometries for T-duality. In
contrast, supersymmetry methods can only be used to construct supersymmetric
solutions, but, in general, they provide very general recipes that permit the
construction of very general families of supersymmetric solutions such as all
supersymmetric black holes of a given Superstring effective field theory (in
the end, a supergravity theory).  In Superstring Theory, supersymmetric
solutions often describe the fields generated by non-perturbative extended
objects such as D$p$-branes and provide a way to learn more about
them. Supersymmetric compactification backgrounds are an essential ingredient
of many superstring phenomenological models as well. But supersymmetric
solutions are also interesting in their own right because, often, they
involve structures and enjoy properties of great physical and mathematical
relevance. All this justifies the great deal of effort employed in the
characterization and classification of all supersymmetric solutions of
Superstring Theory via the characterization and classification of all
supersymmetric solutions of all supergravity theories.

This effort started with the pioneering work of Gibbons, Hull and Tod
\cite{Gibbons:1982fy,Tod:1983pm} in pure $\mathcal{N}=2,d=4$
supergravity. This theory is just the simplest of the very rich family of
$\mathcal{N}=2,d=4$ supergravities which have different matter contents
(vector multiplets and hypermultiplets) and couplings (some of them associated
to gaugings of their global symmetries). All of them have been studied from
this point of view in a long series of papers
\cite{Caldarelli:2003pb,Meessen:2006tu,Huebscher:2006mr,Cacciatori:2008ek,Hubscher:2008yz,Klemm:2009uw,Klemm:2010mc,Meessen:2012sr}
of increasing complexity using the ``spinor bilinear'' method of
Ref.~\cite{Gauntlett:2002nw}, which we will also use and explain in this
paper. The most general case, considered in Ref.~\cite{Meessen:2012sr} has
only been solved for the ``timelike'' case and the ``null'' supersymmetric
solutions of theories with non-Abelian gaugings still have to be
characterized. The supersymmetric solutions (both timelike and null) of the
pure $\mathcal{N}=4,d=4$ theory have also been characterized in
Ref.~\cite{Tod:1995jf,Bellorin:2005zc}, but neither the matter-coupled nor
gauged theories have been studied.\footnote{It is believed that the
  supersymmetric solutions of these, and other supergravity theories are in
  one to one correspondence to those of their $\mathcal{N}=2$ truncations,
  although this has not been formally proven.} Finally, since it is possible
to treat all 4-dimensional supergravities with vector multiplets in a
unified form, all their timelike supersymmetric solutions were characterized
in a unified form in \cite{Meessen:2010fh}. The null case and the gauged
theories remain to be studied.

In $d=5$ dimensions the situation in $\mathcal{N}=1$
theories\footnote{This means 8 supercharges. Sometimes they are
  referred to as $\mathcal{N}=2,d=5$ supergravities in the literature
  although the minimal spinor in 5 dimensions has 8 components because
  of their relation with $\mathcal{N}=2,d=4$ theories.} is better
because all  timelike and null solutions have been characterized
with the most general matter content and couplings in
Refs.~\cite{Gauntlett:2002nw,Gauntlett:2003fk,Gutowski:2004yv,Gauntlett:2004qy,Gutowski:2005id,Bellorin:2006yr,Bellorin:2007yp,Bellorin:2008we}. The
$\mathcal{N}>1,d=5$ theories have not been studied systematically.

In the case of $\mathcal{N}=(1,0),d=6$ supergravity, all supersymmetric
solutions have been characterized systematically in
Refs.~\cite{Gutowski:2003rg,Cariglia:2004kk,Jong:2006za,Akyol:2010iz,Lam:2018jln,Cano:2019gqm},
but those of the rest of the 6-dimensional supergravity theories have not.

For the sake of making this short review of what has been accomplished in this
field of research in $d\leq 6$ complete, let us also mention the work done in
maximal and half-maximal $d=3$ supergravities in
Refs.~\cite{Deger:2010rb,deBoer:2014iba,Colgain:2015mta,Deger:2015tra} and
also $\mathcal{N}=1,d=4$ supergravity in
Refs.~\cite{Gran:2008vx,Ortin:2008wj}. For a review on this topic and
additional references on related work, see
Refs.~\cite{Ortin:2015hya,Gran:2018ijr}.

In dimensions higher than six there are only supergravities with 16 or 32
supercharges. Many of them can be obtained via dimensional reduction from the
10- and 11-dimensional theories and, therefore, most of the work has been
focused directly on these. Pure $\mathcal{N}=1,d=10$ supergravity can be
viewed as the effective field theory of the Heterotic or the Type~I
Superstrings (depending on the stringy interpretation of the supergravity
fields) at lowest order in an expansion in terms of the Regge slope parameter
$\alpha'$. One of the most important features of these theories is the
presence of massless gauge vectors in their spectra. These occur at first
order in $\alpha'$ in the effective action, but there is no problem to
accommodate them in $\mathcal{N}=1,d=10$ supergravity as vector
supermultiplets \cite{Bergshoeff:1981um,Chapline:1982ww}. However, at this
order in $\alpha'$, the Heterotic Superstring effective action contains more
terms which can only be accommodated in $\mathcal{N}=1,d=10$ supergravity if
more terms are also included to preserve invariance under supersymmetry
transformations at a given order
\cite{Bergshoeff:1988nn,Bergshoeff:1989de}. The additional terms are of higher
order in derivatives as well, which leads to very complicated equations of
motion.

From the heterotic superstring effective action point of view, the neglect
of the $\alpha'$ corrections has to be justified in each particular
solution. In general, this imposes constraints on the charges of the solutions
or signals (typically high-curvature) regions of the solutions which cannot be
trusted as good string theory solutions because they are bound to be modified
once the neglected terms in the theory are reconsidered. For this reason, from
the Superstring Theory point of view, it is important to take into account
these possible complicated $\alpha'$ corrections in the analysis from the
onset.

However, even though the corrections to the equations of motion are very
involved, it so happens that the Killing Spinor Equations (KSEs) only get
``implicit'' first-order $\alpha'$ corrections and can be analyzed at zeroth
or first order at the same time. For these reasons, in the first systematic
studies of supersymmetric solutions
Refs.~\cite{Gran:2005wf,Gran:2007fu,Gran:2007kh} the KSEs were solved at
zeroth/first order but involving only the zeroth-order equations of
motion.\footnote{Eqs.~(2.1) of Ref.~\cite{Gran:2005wf}. The first $\alpha'$
  corrections to the equations of motion have been considered more recently in
  \cite{Gran:2018ijr}. See Section~\ref{sec-KSIs} for a discussion on their
  r\^ole in the KSEs' integrability conditions.}

In all these works, the so-called \textit{spinorial geometry} method was used
\cite{Gillard:2004xq}. In this method many computations are streamlined by the
use of a privileged basis of spinors, and, correspondingly, of gamma
matrices. The downside of this approach is that the results obtained
concerning spinors are written in that particular basis and, sometimes, a
basis-independent form could be more desirable for some purposes and could
also give further insights into the structure and interpretation of the
supersymmetric solutions, as we will discuss below. 

Our goal in this paper is to repeat part of the analysis carried out in
Refs.~\cite{Gran:2005wf,Gran:2007fu,Gran:2007kh} using the spinor bilinear
formalism without making any choice of spinor basis, to first order in
$\alpha'$. Thus, we expect to obtain

\begin{enumerate}
\item The algebra of spinor bilinears in an arbitrary basis (see
  Section~\ref{sec-d10bilinearalgebra}). This algebra includes the relations
  satisfied by the Spin(7) structure 4-form $\Omega_{abcd}$.
\item The form of the Killing spinors in a general spinorial basis. In
  particular, we will obtain the supersymmetry projectors using  arbitrary
  gamma matrices (Section~\ref{sec-projectors}). The form of the projector
  Eq.~(\ref{eq:Pi-spinorprojector}) suggests that the solutions with only one
  Killing spinor can be viewed as a multiple intersection of S5-branes, with
  volume form of the 4-dimensional transverse space of each of them entering
  the Spin(7) structure 4-form $\Omega_{abcd}$. This projector can be compared
  with Eq.~(3.10) of Ref.~\cite{Gran:2005wf}. They are equivalent, but the
  later is expressed in a particular basis and its form will change under a
  general change of basis while the form of the projector
  Eq.~(\ref{eq:Pi-spinorprojector}) will not. Furthermore, the intersection
  interpretation is lost in the simplified form.
\item The necessary and sufficient conditions that a supersymmetric field
  configuration has to satisfy, originally found in
  Refs.~\cite{Gran:2005wf,Gran:2007fu,Gran:2007kh}. These will be obtained in
  essentially the same form.
\item The relations that hold between the first-order in $\alpha'$-corrected
  equations of motion when they are evaluated on supersymmetric
  configurations, via the Killing Spinor Identities (see below). 
  Relations of this kind have been derived recently as integrability conditions of the
  Killing Spinor Equations in Ref.~\cite{Gran:2018ijr}. We discuss the
  relation with ours in Section~\ref{sec-KSIs}.
\end{enumerate}

We are just interested in the general characterization of the supersymmetric
solutions (those admitting (at least) one Killing spinor), and we will not try
to study case by case, what happens when the solution admits 2 or more Killing
spinors, as it has been done in full detail in
Refs.~\cite{Gran:2005wf,Gran:2007fu,Gran:2007kh}.

There are several important difficulties which prevent us from making this
analysis at higher orders in $\alpha'$:

\begin{enumerate}

\item Not all $\alpha'$ corrections to the action and supersymmetry
  transformation rules are known. In Ref.~\cite{Bergshoeff:1989de}, which we
  will use here, they were determined to cubic order in $\alpha'$ (quartic
  order in curvatures) by supersymmetrizing the first-order terms (specially
  the Chern-Simons terms in the Kalb-Ramond 3-form field strength) which had
  been found by other means. This supersymmetrization leads to a recursive
  procedure for introducing the terms of next order in $\alpha'$ in the
  Kalb-Ramond 3-form field strength $H$. $H$ occurs in the action and in the
  supersymmetry transformation rules, and part of the $\alpha'$ corrections in
  them are introduced implicitly through $H$. However, apart from these, there
  are other $\alpha'$ corrections of increasing complexity both in the action
  and in the supersymmetry transformation rules and this will force us to work
  only at first order in $\alpha'$ and analyze the KSEs only up to that
  order. It has been suggested that higher order corrections to the KSEs could
  be absorbed into redefinitions of the torsionful spin connection or the
  Kalb-Ramond 3-form field strength which would preserve their first-order
  form, though, \cite{Metsaev:1987zx,Bergshoeff:1989de}. If this was proven,
  the first-order analysis would be sufficient.

\item As we have just said, many (probably most) of the higher-order
  terms in the action cannot be constructed by recursion and they are
  only known explicitly up to cubic order. The equations of
  motion\footnote{We only consider purely bosonic configurations and,
    therefore, we always refer, implicitly, to the equations of motion
    of the bosonic fields of the theory.}  have a very large number of
  complicated terms. At first order, though, it is known that many of
  them can be ignored because they are proportional to the
  zeroth-order equations of motion. It is not known if similar
  simplifications take place at higher orders and, if one wants to
  find explicit solutions at higher orders one will have to deal with
  very complicated equations.
  
\item In most of the literature on the characterization or classification of
  supersymmetric solutions the following fact is used: the integrability
  conditions of the KSEs (adequately treated) are combinations of the
  off-shell equations of motion. This leads to non-trivial relations between
  them which simplify the problem of finding supersymmetric solutions, because
  only a small number of equations of motion are independent and have to be
  solved. At higher orders in $\alpha'$, computing the integrability
  conditions of the KSEs and recognizing in them the complicated higher-order
  equations of motion can be very difficult.

  The Killing Spinor Identities (KSIs) derived in Ref.~\cite{Kallosh:1993wx}
  offer an alternative path, because, based on the invariance of the theory
  under local supersymmetry transformations, they yield the same relations
  between equations of motion\footnote{As we will explain in
    Section~\ref{sec-EOM}, the equations of motion usually considered in the
    literature are combinations of those one obtains by direct variation of
    the action where terms proportional to the lowest-order equations of
    motion are eliminated. In order to use the KSIs, we need to take into
    account these facts.} even if the form of the latter is not known
  explicitly \cite{Bellorin:2005hy}. The main ingredients are the
  supersymmetry transformations of the bosonic fields of the theory, which,
  usually, are much simpler than those of the fermions.
 
\item However, even though, in most cases, the supersymmetry transformations
  of the bosonic fields are not modified when the supergravity theory is
  gauged or deformed in other ways, in this case the Kalb-Ramond 3-form field
  strength is modified at first order in $\alpha'$ by the addition of the
  Lorentz- and Yang-Mills-Chern-Simons terms that modify the gauge
  transformations of the Kalb-Ramond 2-form. The supersymmetry transformations
  of this field acquire additional terms related to these gauge
  transformations which results in a non-trivial modification of the KSIs at
  first order in $\alpha'$ we have to deal with.
   
\item The main drawback of the KSIs is that, by construction, they never
  include the Bianchi identities satisfied by the field strengths in the
  relations obtained. The Bianchi identity of the Kalb-Ramond 3-form is
  typically one of the key equations to be solved, though.  This deficiency of
  the KSI approach can be overcome by including the 6-form dual of the
  Kalb-Ramond 2-form in the derivation of the KSIs, as we will show in
  Section~\ref{sec-KSIs}. The equation of motion of the 6-form is, by
  definition, the Bianchi identity of the Kalb-Ramond 3-form field
  strength. We are not including the Bianchi identity of the Yang-Mills gauge
  field strength because in order to write it one needs to know the gauge
  connection, which completely determines and trivializes the Bianchi
  identity.

\end{enumerate}

After considering all these difficulties and the solutions found for some of
them, it is clear that, being conservative, we will have to content ourselves
with carrying our program to first order in $\alpha'$ only: we will determine
necessary and sufficient conditions for unbroken supersymmetry valid to
$\mathcal{O}(\alpha')$ and relations between equations of motion evaluated on
supersymmetric configurations valid to the same order in $\alpha'$. We will
not be able to simplify the independent equations of motion beyond
$\mathcal{O}(\alpha')$, either, because it is not clear if they can be
simplified.

This paper is organized as follows: we start by reviewing
Section~\ref{sec-heteroticalpha} the bosonic Heterotic Superstring effective
action, equations of motion and supersymmetry transformation
rules. Appendix~\ref{sec-d10spinorsetc} contains our conventions for gamma
matrices and spinors, the definitions of the spinor bilinears used in the main
text and the computation of the algebra satisfied by these bilinears using the
Fierz identities in an arbitrary basis which we are going to use in
Section~\ref{sec-d10hetconf}.\footnote{We have learned, after doing this
  calculation, that there is a much more efficient way of computing the
  bilinear algebra~\cite{Cortes:2019xmk}.} As a byproduct we will derive the algebra
of the Spin(7) structure 4-form which is always present in the bilinear
algebra.  In Section~\ref{sec-d10hetconf} we will determine the necessary
conditions for a field configuration to admit one Killing spinor (summarized
in~\ref{sec-d10summary}) and we will show that they are also sufficient by
solving explicitly the KSEs. We will make use of the spinor projector given
explicitly in terms of the Spin(7) structure in
Eq.~(\ref{eq:Pi-spinorprojector}). In Section~\ref{sec-d10hetsol} we determine
which independent equations have to be imposed on the supersymmetric field
configurations determined in the previous section to ensure that they are
solutions of all equations of motion of the theory. We explain how the
equations of motion are obtained and simplified at first order in $\alpha'$
and the derivation of the KSIs involving also the Bianchi identity of the
Kalb-Ramond 3-form field strength. Finally, Section~\ref{sec-discussion}
contains a brief discussion of the results obtained and their spinoffs.

\section{The Heterotic Superstring effective action}
\label{sec-heteroticalpha}

In this section we are going to review the Heterotic Superstring effective
action was given in Ref.~\cite{Bergshoeff:1989de}, where it was constructed up
to cubic order in $\alpha'$ (quartic in derivatives) by demanding invariance
of the action under supersymmetry (to that order). Here we will use the
conventions of Ref.~\cite{Ortin:2015hya}.\footnote{The relation with the
  fields in Ref.~\cite{Bergshoeff:1989de} is as follows: the metric and gauged
  fields can be identified; the Kalb-Ramond fields are related by
  $B_{{\rm BdR}\, \mu\nu}=\tfrac{1}{\sqrt{2}}B_{\mu\nu}$ and their field
  strengths by
  $H_{{\rm BdR}\, \mu\nu\rho}=\tfrac{1}{3\sqrt{2}}H_{\mu\nu\rho}$. The dilaton
  fields are related by $\phi_{\rm BdR}=e^{2\phi/3}$. The gravitino and
  dilatino are related by $\psi_{\rm BdR}{}_{\mu}=\sqrt{2}\psi_{\mu}$,
  $\lambda_{\rm BdR}=-\tfrac{1}{2}\lambda$ while the gaugini are related by
  $\chi^{A}_{\rm BdR}=\sqrt{2}\chi^{A}$. The relation between the 6-form dual
  of the Kalb-Ramond 2-form in
  Refs.~\cite{Bergshoeff:1990ax,Bergshoeff:1990hh},
  $A_{\alpha_{1}\cdots\alpha_{6}}$ by the same authors and our 6-form
  $\tilde{B}_{\alpha_{1}\cdots\alpha_{6}}$ is
  $A_{\alpha_{1}\cdots\alpha_{6}} = \tfrac{1}{2\sqrt{2}\cdot
    6!}\tilde{B}_{\alpha_{1}\cdots\alpha_{6}}$. Finally, the supersymmetry
  parameters are related by $\epsilon_{\rm BdR}=\sqrt{2}\epsilon$ and the
  parameters $\alpha$ and $\beta$ are both equal to $\alpha'/4$.}

We start by defining recursively the 3-form field strength of the Kalb-Ramond
2-form $B$. The zeroth-order, it is given by 

\begin{equation}
H^{(0)} \equiv dB\, .
\end{equation}

\noindent
Using it as torsion, one can define the zeroth-order torsionful spin (or
Lorentz) connections

\begin{equation}
{\Omega}^{(0)}_{(\pm)}{}^{{a}}{}_{{b}} 
=
{\omega}^{{a}}{}_{{b}}
\pm
\tfrac{1}{2}{H}^{(0)}_{{\mu}}{}^{{a}}{}_{{b}}dx^{{\mu}}\, ,
\end{equation}

\noindent
where $\omega^{a}{}_{b}$ is the (torsionless, metric-compatible)
Levi-Civita spin connection 1-form.

The curvature 2-forms of these connections and the (Lorentz-) Chern-Simons
3-forms are defined as\footnote{The same definition applies to other spin
  connections with the same indices; in particular, for the curvature 2-form
  of the Levi-Civita connection (the Riemann tensor).}

\begin{eqnarray}
{R}^{(0)}_{(\pm)}{}^{{a}}{}_{{b}}
& = & 
d {\Omega}^{(0)}_{(\pm)}{}^{{a}}{}_{{b}}
- {\Omega}^{(0)}_{(\pm)}{}^{{a}}{}_{{c}}
\wedge  
{\Omega}^{(0)}_{(\pm)}{}^{{c}}{}_{{b}}\, ,
\\
& & \nonumber \\
{\omega}^{{\rm L}\, (0)}_{(\pm)}
& = &  
d{\Omega}^{ (0)}_{(\pm)}{}^{{a}}{}_{{b}} \wedge 
{\Omega}^{ (0)}_{(\pm)}{}^{{b}}{}_{{a}} 
-\tfrac{2}{3}
{\Omega}^{ (0)}_{(\pm)}{}^{{a}}{}_{{b}} \wedge 
{\Omega}^{ (0)}_{(\pm)}{}^{{b}}{}_{{c}} \wedge
{\Omega}^{ (0)}_{(\pm)}{}^{{c}}{}_{{a}}\, .  
\end{eqnarray}

We will denote the gauge field 1-form by $A^{A}$, where the indices
$A,B,C,\ldots$ take values in the Lie algebra of the gauge group. The
(Yang-Mills) gauge field strength and Chern-Simons 3-form are defined by

\begin{eqnarray}
{F}^{A}
& = & 
d{A}^{A}+\tfrac{1}{2} f_{BC}{}^{A}{A}^{B}\wedge{A}^{C}\, , 
\\
& & \nonumber \\
{\omega}^{\rm YM}
& = & 
dA_{A}\wedge {A}^{A}
+\tfrac{1}{3}f_{ABC}{A}^{A}\wedge{A}^{B}\wedge{A}^{C}\, ,
\end{eqnarray}

\noindent
where the Killing metric of the gauge group's Lie algebra in the relevant
representation, $K_{AB}$, assumed to be invertible and positive definite, has
been used to lower the index of the structure constants
$f_{ABC}\equiv f_{AB}{}^{D}K_{DB}$ and of the indices of the gauge fields
$A_{A}\equiv K_{AB}A^{B}$.

Then, using the Yang-Mills- and zeroth-order Lorentz-Chern-Simons
3-forms,\footnote{Only $\Omega_{(-)}$ occurs in $H$.} the
first-order the Kalb-Ramond 3-form field strength is defined to be

\begin{equation}
\label{eq:H1def}
  H^{(1)}
= 
d{B}
+\frac{\alpha'}{4}\left({\omega}^{\rm YM}
+{\omega}^{{\rm L}\, (0)}_{(-)}\right)\, ,    
\end{equation}

\noindent
and using it as torsion, we obtain the first-order torsionful spin connections

\begin{equation}
{\Omega}^{(1)}_{(\pm)}{}^{{a}}{}_{{b}} 
=
{\omega}^{{a}}{}_{{b}}
\pm
\tfrac{1}{2}{H}^{(1)}_{{\mu}}{}^{{a}}{}_{{b}}dx^{{\mu}}\, ,
\end{equation}

\noindent
and their curvatures and Lorentz-Chern-Simons terms
${R}^{(1)}_{(\pm)}{}^{{a}}{}_{{b}},{\omega}^{{\rm L}\, (1)}_{(\pm)}$ are
obtained by plugging them into the above definitions. Then, the second-order
Kalb-Ramond field strength is defined as

\begin{equation}
H^{(2)}
= 
d{B}
+\frac{\alpha'}{4}\left({\omega}^{\rm YM}
  +{\omega}^{{\rm L}\, (1)}_{(-)}\right)\, ,\,\,\,\,
\ldots
\,\,\,\,\,\,
H^{(n)}
= 
d{B}
+\frac{\alpha'}{4}\left({\omega}^{\rm YM}
  +{\omega}^{{\rm L}\, (n-1)}_{(-)}\right)\, ,
\end{equation}

\noindent
etc.

For many practical purposes it is advantageous to work with general $H$ and
$\Omega_{(\pm)}$ and then restrict them to a given order when needed. This
will allow us to work with the Killing spinor equations at an arbitrary order
in $\alpha'$, for instance, because the only $\alpha'$ corrections are
contained in the definitions of $H$ and $\Omega_{(+)}$. In the action, apart
from the $\alpha'$ corrections implicit in the definitions of $H$ and
$\Omega_{(-)}$, there are additional terms of higher order in curvatures that
have to be included explicitly and which are known only to cubic order in
$\alpha'$ \cite{Bergshoeff:1989de}.\footnote{And not completely: only the
  quartic terms that follow from the supersymmetrization of the Chern-Simons
  terms in $H$ were determined, but there may be more
  \cite{Gross:1986iv,Grisaru:1986dk,Grisaru:1986kw}. } It is understood that
all terms above a certain order in $\alpha'$ have to be ignored. Thus, with
this understanding, we will omit the upper indices $(n)$ from now on.

It is convenient to define an affine torsionful connection
$\Gamma^{(+)}{}_{\mu\nu}{}^{\rho}$ via the Vielbein postulate

\begin{equation}
  \nabla^{(+)}{}_{\mu}e^{a}{}_{\nu}
  =
  \partial_{\mu}e^{a}{}_{\nu}
  -\Omega^{(\pm)}{}_{\mu}{}^{a}{}_{b}e^{b}{}_{\nu}
  -\Gamma^{(\pm)}{}_{\mu\nu}{}^{\rho}e^{a}{}_{\rho}
  =
  0\, .
\end{equation}

\noindent
Solving the above equation one finds that it is given by

\begin{equation}
  \label{eq:torsionfulaffineconnection}
  \Gamma^{(\pm)}{}_{\mu\nu}{}^{\rho}
  =
  \left\{\!\!
    \begin{array}{c}
      _{\rho} \\ _{\mu\nu} \\
    \end{array}
    \!\!
  \right\}
  \pm
  \tfrac{1}{2} H_{\mu\nu}{}^{\rho}\, ,
\end{equation}

\noindent
where
$\left\{\!\!  \begin{array}{c} _{\rho} \\ _{\mu\nu} \\ \end{array}
  \!\!  \right\}$ stands for the Christoffel symbols of the metric
$g_{\mu\nu}=\eta_{ab}e^{a}{}_{\mu}e^{b}{}_{\nu}$.

It is also convenient to use the so-called ``$T$-tensors'' associated to the
$\alpha'$ corrections in the equations of motion and in the Bianchi identity
of the Kalb-Ramond 3-form field strength. They are defined by 

\begin{equation}
\label{eq:Ttensors}
\begin{array}{rcl}
{T}^{(4)}
& \equiv &
\dfrac{\alpha'}{4}\left[
{F}_{A}\wedge{F}^{A}
+
{R}_{(-)}{}^{{a}}{}_{{b}}
\wedge
{R}_{(-)}{}^{{b}}{}_{{a}}
\right]\, ,
\\
& & \\ 
{T}^{(2)}{}_{{\mu}{\nu}}
& \equiv &
\dfrac{\alpha'}{4}\left[
{F}_{A}{}_{{\mu}{\rho}}{F}^{A}{}_{{\nu}}{}^{{\rho}} 
+
{R}_{(-)\, {\mu}{\rho}}{}^{{a}}{}_{{b}}
{R}_{(-)\, {\nu}}{}^{{\rho}\,  {b}}{}_{{a}}
\right]\, ,
\\
& & \\    
{T}^{(0)}
& \equiv &
{T}^{(2)\,\mu}{}_{{\mu}}\, .
\\
\end{array}
\end{equation}

The Heterotic Superstring effective action, written in the string frame to
cubic order in $\alpha'$ in terms of the objects we have just defined tales
the form

\begin{equation}
\label{heterotic}
S
=
\frac{g_{s}^{2}}{16\pi G_{N}^{(10)}}
\int d^{10}x\sqrt{|{g}|}\, 
e^{-2{\phi}}\, 
\left\{
{R} 
-4(\partial{\phi})^{2}
+\frac{1}{12}{H}^{2}
-\frac{1}{2}T^{(0)}
-\frac{\alpha'}{48} (T^{(4)})^{2}
-\frac{\alpha'}{4} (T^{(2)})^{2}
\right\}\, ,
\end{equation}

\noindent
where $R$ is the Ricci scalar of the string-frame metric $g_{\mu\nu}$, $\phi$
is the dilaton field and its vacuum expectation value of $e^{\phi}$ is the
Heterotic Superstring coupling constant $g_{s}$ and where $G_{N}^{(10)}$ is
the 10-dimensional Newton constant.

Observe that, to have all $\mathcal{O}(\alpha'^{3})$ terms in the action, we
have to use $H^{(3)}$ and $\Omega^{(3)}_{(-)}$, disregarding all terms of
higher order that arise from $H^{2}$ etc. in the above action. As explained in
the Introduction, though, we will only work at first order in $\alpha'$.

Finally, let us consider the supersymmetry transformation rules of the bosons
and of the fermionic fields (gravitino $\psi_{\mu}$, dilatino $\lambda$ and
gaugini $\chi^{A}$)\footnote{All fermions and the supersymmetry parameter
  $\epsilon$ are Majorana-Weyl spinors and $\epsilon$ has positive chirality
  in the conventions given in Appendix~\ref{sec-d10spinorsetc}.} for vanishing
fermions and to first order in $\alpha'$. They can be written, respectively,
as follows:

\begin{eqnarray}
  \label{eq:susyrule1}
  \delta_{\epsilon} e^{a}{}_{\mu}
  & = &
        \bar{\epsilon}\Gamma^{a}\psi_{\mu}\, ,
  \\
  & & \nonumber \\
  \label{eq:susyruleKR}
  \delta_{\epsilon} B_{\mu\nu}
  & = &
        2\bar{\epsilon}\Gamma_{[\mu}\psi_{\nu]} +\frac{\alpha'}{2}
        \left\{ A_{A\, [\mu|}\delta_{\epsilon}A^{A}{}_{|\nu]}
        +\Omega_{(-)\, [\mu|}{}^{a}{}_{b}
        \delta_{\epsilon}\Omega_{(-)\, |\nu]}{}^{b}{}_{a}
        \right\}\, ,
  \\
  & & \nonumber \\
  \label{eq:susytrans6-form}
  \delta_{\epsilon} \tilde{B}_{\mu_{1}\cdots\mu_{6}}
  & = &
        6e^{-2\phi}\bar{\epsilon}\Gamma_{[\mu_{1}\cdots\mu_{5}}\left[\psi_{\mu_{6}]}
        -\tfrac{1}{6}\Gamma_{\mu_{6}]}\lambda\right]\, ,
  \\
  & & \nonumber \\
  \delta_{\epsilon}\phi
  & = &
        \tfrac{1}{2}\bar{\epsilon}\lambda\, ,
          \\
  & & \nonumber \\
  \label{eq:susyrule5}
  \alpha'\delta_{\epsilon}A^{A}{}_{\mu}
  & = &
        \alpha'\bar{\epsilon}\Gamma_{\mu}\chi^{A}\, ,
\end{eqnarray}

\noindent
where, at this order,

\begin{equation}
  \delta_{\epsilon}\Omega_{(-)\, \mu}{}^{ab}
  =
  \bar{\epsilon}\Gamma_{\mu}\psi^{ab}\, ,
\end{equation}

\noindent
with the gravitino field strength $\psi_{\mu\nu}$ defined as

\begin{equation}
  \psi_{\mu\nu}
  \equiv
  2\mathcal{D}^{(+)}_{[\mu}\psi_{\nu]}
  \equiv
  2\partial_{[\mu}\psi_{\nu]}
  -\tfrac{1}{2}\Omega_{(+)\, [\mu|}{}^{ab}\Gamma_{ab}\psi_{|\nu]}\, ,
\end{equation}

\noindent
and

\begin{eqnarray}
  \delta_{\epsilon} \psi_{a}
  & = &
        \nabla^{(+)}{}_{a}\,  \epsilon
        \equiv
        \left(\partial_{a}-\tfrac{1}{4} \Omega^{(+)}{}_{a \, bc} \Gamma^{bc}\right)
        \epsilon \, ,
  \\
  \nonumber \\
  \delta_{\epsilon} \lambda
  & = &
        \bigg( \partial_{a} \phi \Gamma^{a}
        -\tfrac{1}{12} H_{abc} \Gamma^{abc} \bigg) \epsilon \, ,
  \\
  \nonumber \\
  \alpha'\delta_{\epsilon} \chi^{A}
  & = &
        - \tfrac{1}{4} \alpha' F^{A}{}_{ab} \Gamma^{ab} \epsilon\, . 
\end{eqnarray}

\section{Supersymmetric configurations}
\label{sec-d10hetconf}

The (purely bosonic) supersymmetric field configurations of this
theory are those for which the Killing Spinor Equations (KSEs)
$\delta_{\epsilon} \psi_{a}= \delta_{\epsilon} \lambda = 
\delta_{\epsilon} \chi^{A} =0$ admit at least one solution called
Killing spinor that we will denote by $\epsilon$. Thus, if
$e^{a}{}_{\mu},B_{\mu\nu},\phi$ describe a supersymmetric field
configuration of this theory, there is an $\epsilon$ satisfying the
KSEs

\begin{eqnarray}
\label{KSE_gravitino}
        \nabla^{(+)}{}_{a}\, \epsilon
  & = &
        0\, ,
  \\
  \nonumber \\
\label{KSE_dilatino}
        \bigg( \partial_{a} \phi \Gamma^{a}
        -\tfrac{1}{12} H_{abc} \Gamma^{abc} \bigg) \epsilon 
  & = &
        0\, ,
  \\
  \nonumber \\
\label{KSE_gaugino}
F^{A}{}_{ab} \Gamma^{ab} \epsilon
  & = &
        0\, . 
\end{eqnarray}
  
When the spinor bilinears $\ell_{a}$ and $W_{a_{1}\cdots a_{5}}$
defined in Appendix~\ref{sec-d10bilinears}, are constructed with the
Killing spinor $\epsilon$ that satisfies the above equations, they
must satisfy certain some other equations apart from the algebraic
relations found in Appendix~\ref{sec-d10bilinearalgebra}. In what
follows, we are going to determine those equations and their immediate
consequences, for each KSE.

\subsection{The gravitino KSE}

Using the torsionful affine connection defined in
Eq.~(\ref{eq:torsionfulaffineconnection}), the gravitino KSE
Eq.~(\ref{KSE_gravitino}) immediately leads to these two differential
equations:

\begin{eqnarray}
\label{grav_1}
  \nabla^{(+)}_{a} \ell_{b} & = &  0\, ,
  \\
  \nonumber \\
\label{grav_2}
\nabla^{(+)}_{a} W_{b_{1} \cdots b_{5}}  & = &  0\, . 
\end{eqnarray}

Using Eq.~(\ref{eq:WlO}), these two equations lead to another equation
for the 4-form $\Omega_{a_{1}\cdots a_{4}}$

\begin{equation}
\label{grav_3}
\ell_{b_{1}}\nabla^{(+)}_{a} \Omega_{b_{2} \cdots b_{4}}  =   0\, ,
\end{equation}

\noindent
where we are using the same convention as in the appendix: all indices
with the same Latin letter (here $b_{1}\cdots b_{5}$) are assumed to
be fully antisymmetrized.

The symmetric and antisymmetric parts of Eq.~(\ref{grav_1}) indicate
that the null vector $\ell_{a}$ is a Killing vector

\begin{equation}
  \label{grav_4}
  \nabla_{(a}\ell_{b)}=0\, ,  
\end{equation}

\noindent
($\nabla$ is the standard Levi-Civita connection) and that

\begin{equation}
  \label{grav_5-1}
2\nabla_{[a}\ell_{b]} = \ell^{c}H_{cab}\, ,  
\end{equation}

\noindent
or, in the language of differential forms

\begin{equation}
  \label{grav_5-2}
i_{\ell}H = d\ell\, .
\end{equation}

It is customary to introduce an auxiliary null vector $n$, dual to $\ell$

\begin{equation}
  \label{eq:ndefinition}
  n^{2}=0\, ,
  \hspace{1cm}
  n^{a}\ell_{a}=1\, .
\end{equation}

Then, we can use the 1-forms 

\begin{equation}
\ell_{\mu}dx^{\mu}\equiv e^{+} \, , \qquad\qquad 
n_{\mu}dx^{\mu}\equiv e^{-}
\end{equation}
as the first two elements of a Vielbein basis $\{e^{+},e^{-},e^{m}\}$ $m=1,\ldots,8$ in which the metric takes the form

\begin{equation}
ds^{2} = e^{+}\otimes e^{-} +e^{-}\otimes e^{+} -\delta_{mn}e^{m}\otimes e^{n}\, .
\end{equation}

Eq.~(\ref{grav_5-2}) can be interpreted as the $+$ component of the
first Cartan structure equations
($de^{a}=\omega^{a}{}_{b}\wedge e^{b}$ in our conventions) and from it
we find that 

\begin{equation}
  \label{eq:omega[ab]-0}
\Omega^{(+)}{}_{[ab]-}=0\, .  
\end{equation}

On the other hand, from Eq.~(\ref{grav_3}) we get these two equations
in the above basis:

\begin{subequations}
  \begin{align}
    \label{grav_3-1}
  \nabla^{(+)}{}_{a}\Omega_{m_{1}\cdots m_{4}}
  & =
    0\, ,
  \\
    \nonumber \\
        \label{grav_3-2}
\nabla^{(+)}{}_{a}\Omega_{m_{1}m_{2} m_{3} -}
  & =
    0\, .  
\end{align}
\end{subequations}

\noindent
Since, in this basis, the 4-form $\Omega$'s only non-vanishing
components are those with transverse indices $m,n,p,\ldots$, 
Eq.~(\ref{grav_3-2}) implies that

\begin{equation}
  \label{eq:omega+am-=0}
\Omega^{(+)}{}_{am-}=0\,.  
\end{equation}

In order to analyze Eq.~(\ref{grav_3-1}) we need a more detailed choice of
coordinates that we are going to make next.

\subsubsection{The metric}

All metrics characterized by the existence of a null, generically not
covariantly-constant nor hypersurface-orthogonal, Killing vector can be
written in a common way (see, \textit{e.g.}~Ref.~\cite{Cano:2019gqm}: first of
all, we introduce null coordinates $v,u$ through

\begin{subequations}
  \begin{align}
    \partial_{-}
    & \equiv 
               \partial_{v}\, ,
    \\
    \nonumber \\
    e^{+} & \equiv
            f(du+\beta)\, ,
  \end{align}
\end{subequations}

\noindent
where $\beta=\beta_{\underline{m}}dx^{m}$, $m=1,\cdots,8$ is a 1-form
in the 8-dimensional space orthogonal to $e^{+}$, $f$ is a scalar
function and both $f$ and $\beta$ are independent of $v$ (but, in
general, not of $u$ nor of the remaining coordinates $x^{m}$).

Next, we write $e^{-}$ in the form

\begin{equation}
  e^{-} \equiv dv +Kdu +\omega\, ,  
\end{equation}

\noindent
where $\omega=\omega_{\underline{m}}dx^{m}$ is another 1-form in the
transverse 8-dimensional space and $K$ is a scalar function which are
also independent of $v$.

Choosing the Vielbeins $e^{m}$ to only have non-vanishing
$v$-independent components in the transverse directions,
$e^{m}=e^{m}{}_{\underline{n}}dx^{n}$, the metric takes the form

\begin{equation}
  \label{eq:metric}
ds^{2}
=
2f(du+\beta)(dv+Kdu+\omega)
-h_{\underline{m}\underline{n}}dx^{m} dx^{n}\, ,
\end{equation}

\noindent
where the metric in the transverse space is given by

\begin{equation}
h_{\underline{m} \underline{n}}
= 
\delta_{mn}e^{m}{}_{\underline{m}}e^{n}{}_{\underline{n}}\, .
\end{equation}

It is clear, then, that the transverse components of the spin
connection $\omega_{mnp}$ only depend on the transverse Vielbeins
$e^{m}$ and the $a=m$ components of  Eq.~(\ref{grav_3-1})

\begin{equation}
  \nabla^{(+)}{}_{m}\Omega_{n_{1}\cdots n_{4}}=0\, ,
\end{equation}

\noindent
are those of an equation in transverse space. We can rewrite it as

\begin{equation}
  \nabla_{m}\Omega_{n_{1}\cdots n_{4}}= 2H_{mn_{1}}{}^{p}\Omega_{pn_{2}n_{3} n_{4}}\, ,
\end{equation}

\noindent
and multiplying it by $\Omega_{q}{}^{n_{2}n_{3}n_{4}}$ and using
Eqs.~(\ref{eq:O3O3}) and (\ref{eq:O2O2bis}) we find

\begin{equation}
  \label{eq:HmnpPi}
  H_{mnp}
  =
  -\tfrac{7}{2}H^{(-)}_{mnp}
  +\tfrac{1}{8}\nabla_{[m}\Omega_{n}{}^{s_{1}s_{2}s_{3}}\Omega_{p]\, s_{1}s_{2}s_{3}}\, ,
\end{equation}

\noindent
where we have used the projector acting on 3-forms defined in
Appendix~\ref{sec-projectors}.

The  $a=+$ components of Eq.~(\ref{grav_3-1}), on the other hand, can be
written in the form

\begin{equation}
  \nabla_{+}\Omega_{m_{1}\cdots m_{4}}
  =
  2H_{+m_{1}}{}^{n}\Omega_{nm_{2}m_{3} m_{4}}\, ,
\end{equation}

\noindent
and using the same properties, we find

\begin{equation}
  H^{(-)}_{+mn}
  =
  \tfrac{1}{48}\Omega_{m}{}^{s_{1}s_{2}s_{3}}\nabla_{+}\Omega_{n s_{1}s_{2}
    s_{3}}\, .
\end{equation}

Since, in the coordinates and frame we have chosen,
$\omega_{-mn}=-\omega_{mn-}$, Eq.~(\ref{eq:omega+am-=0}) leads to

\begin{equation}
  \label{eq:omega+-mn=0}
  \Omega^{(+)}{}_{-mn}= H_{-mn}\,,  
\end{equation}

\noindent
and the $a=-$ component of Eq.~(\ref{grav_3-1}) can be written in the form

\begin{equation}
  \partial_{-}\Omega_{m_{1}m_{2}m_{3} m_{4}}
  =
  H_{-m_{1}}{}^{n}\Omega_{nm_{2}m_{3}m_{4}}\,,
\end{equation}

\noindent
and, using the same properties as in the case of the $a=+$ component, we get

\begin{equation}
  \label{eq:H--mn}
  H^{(-)}_{-mn}
  =
  \tfrac{1}{24}\Omega_{m}{}^{s_{1}s_{2}s_{3}}
  \partial_{-}\Omega_{n s_{1}s_{2}s_{3}}\,.
\end{equation}

Finally, observe that the components of the spin connection are
determined by the objects that occur in the metric: the scalar
functions $f,K$, the transverse 1-forms $\omega,\beta$ and the
transverse metric $h$. Via Eqs.~(\ref{eq:omega+am-=0}) they also
determine the $H_{am-}$ components of the Kalb-Ramond field strength.
These components are constrained by the dilatino KSE and the
constraints become constraints on the objects that occur in the
metric.

\subsection{The dilatino KSE}

Multiplying the dilatino KSE by $\bar{\epsilon}$ from the left, we get

\begin{equation}
  \label{dila_1}
  \ell^{\mu}\partial_{\mu}\phi = 0\, ,
  \,\,\,\,\,
  \Rightarrow
  \,\,\,\,\,
  \partial_{v}\phi=0\, .
\end{equation}

If we multiply by $\bar{\epsilon}\Gamma_{ab}$  from the left, we get

\begin{equation}
\label{dila_2}
2\ell_{[a}\partial_{b]}\phi
-\tfrac{1}{12}W_{ab}{}^{cde}H_{cde}
+\tfrac{1}{2}\ell^{c}H_{cab}
 = 
   0\, ,
\end{equation}

\noindent
In terms of the 4-form $\Omega$ we arrive to

\begin{equation}
  e^{+}{}_{[a} \left[2\partial_{b]}\phi
    -\tfrac{1}{6}\Omega_{b]}{}^{cde}H_{cde}\right]
  -\tfrac{1}{4}\Omega_{ab}{}^{cd}H_{cd-}+\tfrac{1}{2}H_{ab-}
  =0\, ,
\end{equation}

The $a=+,m$ components of this equation give a pair of non-trivial
equations in transverse space\footnote{Observe that we could write,
  using Eq.~(\ref{dila_1}) $d\phi -i_{n}d\phi \ell=\tilde{d}\phi$.}

\begin{subequations}
  \begin{align}
      \label{dila_3}
    \tfrac{1}{6}\Omega_{m}{}^{npq}H_{npq}
    +H_{+-m}-2\partial_{m}\phi
    & = 0
      \, ,
    \\
    \nonumber \\
    \label{dila_4}
    H^{(-)}_{qr-}
    & =
      0\, .
  \end{align}
\end{subequations}

The last of these equations, together with Eq.~(\ref{eq:H--mn}) leads to

\begin{equation}
\partial_{-}\Omega_{m_{1}m_{2}m_{3}m_{4}}=0\,.  
\end{equation}

If we multiply the dilatino KSE by 
$\bar{\epsilon}\Gamma_{a_{1}\cdots a_{4}}$ from the left, we get 

\begin{equation}
  \label{dila_5}
  W_{a_{1}\cdots a_{4}}{}^{c}\partial_{c}\phi
  -W_{a_{1}a_{2}a_{3}}{}^{bc}H_{a_{4}bc} +2\ell_{a_{1}}H_{a_{2}a_{3}a_{4}}
  = 0\, .
\end{equation}

Using again Eq.~(\ref{eq:WlO}) we get another pair of equations in
transverse space

\begin{subequations}
  \begin{align}
      \label{dila_6}
    (2\partial_{n}\phi -H_{+-n})\Omega^{n}{}_{m_{1}m_{2}m_{3}}
    +\tfrac{3}{2}H_{m_{1}}{}^{np}\Omega_{m_{2}m_{3}np} -H_{m_{1}m_{2}m_{3}}
    & =
      0\, ,
    \\
    \nonumber \\
      \label{dila_7}
    \Omega_{m_{1}m_{2}m_{3}}{}^{n}H_{m_{4}n-}
    & =
      0\, .
  \end{align}
\end{subequations}

It turns out that these two equations are just combinations of
Eqs.~(\ref{dila_3}) and (\ref{dila_4}): if we multiply
Eq.~(\ref{dila_7}) by the 4-form $\Omega$, contracting just 3 of the 4
free indices and using Eqs.~(\ref{eq:O3O3}) and (\ref{eq:O2O2noantisimplebis})
we obtain Eq.~(\ref{dila_4}). The same happens with
Eq.~(\ref{dila_6}). Therefore, the only three independent equations
one obtains from the dilatino KSE are Eq.~(\ref{dila_1}),
(\ref{dila_3}) and (\ref{dila_4}).

Eq.~(\ref{dila_6}) can be rewritten in the form 

\begin{equation}
  \label{eq:H-mnp}
H^{(-)}_{mnp} = \tfrac{1}{7}(2\partial_{q}\phi -H_{+-q})\Omega^{q}{}_{mnp}\, .
\end{equation}

\noindent
and, combining this result with Eq.~(\ref{eq:HmnpPi}) we can solve for the
components $H_{mnp}$:

\begin{equation}
  \label{eq:Hmnp}
  H_{mnp}
  =
  -\tfrac{1}{2}(2\partial_{q}\phi -H_{+-q})\Omega^{q}{}_{mnp}
  +\tfrac{1}{8} \Omega_{[m}{}^{s_{1}s_{2}s_{3}}
  \nabla_{n}\Omega_{p]s_{1}s_{2}s_{3}}\, .
\end{equation}

\subsection{The gaugino KSE}

Multiplying the gaugino KSE Eq.~(\ref{KSE_gaugino}) by
$\bar{\epsilon}\Gamma_{a}$, $\bar{\epsilon}\Gamma_{abc}$ and
$\bar{\epsilon}\Gamma_{a_{1}\cdots a_{5}}$ from the left, we get

\begin{subequations}
\label{gau_1}
\begin{align}
  \ell^{b}F^{A}{}_{ba}
  & =
    0\, ,
    \,\,\,\,\,
    \Rightarrow
    \,\,\,\,\,
    F^{A}{}_{a-}=0\, ,
  \\
  \nonumber \\
\label{gau_2}
  F^{A\, c_{1}c_{2}}W_{c_{1}c_{2}a_{1}a_{2}a_{3}}
  -6\ell_{a_{1}}F^{A}{}_{a_{2}a_{3}}
  & =
    0\, ,
  \\
  \nonumber \\
  \label{gau_3}
  F^{A}{}_{a_{1}}{}^{c}W_{ca_{2}\cdots a_{4}}
  & =
    0\, .
\end{align}
\end{subequations}

\noindent
respectively.

Using Eq.~(\ref{gau_1}) and the decomposition of $W$ in terms of $\ell$
and $\Omega$ Eqs.~(\ref{gau_2}) and (\ref{gau_3}) lead to

\begin{subequations}
  \begin{align}
\label{gau_2_reduced}
    F^{A}{}_{pq}\Pi^{(-)\, pq}{}_{mn}
    & =
      0\, ,
    \\
    \nonumber \\
    \label{gau_3_reduced}
    \Omega_{m_{1}m_{2}m_{3}}{}^{n}F^{A}{}_{m_{4}n}
    & =
      0\, .
  \end{align}
\end{subequations}

Observe that these two equations for $F^{A}{}_{mn}$ have exactly the
same form as Eqs.~(\ref{dila_4}) and (\ref{dila_7}) for $H_{mn-}$ and,
therefore, they are equivalent by virtue of the properties of the
4-form $\Omega$. The components $F^{A}{}_{m+}$ remain undetermined.

Eq.~(\ref{gau_2_reduced}) is the natural generalization of the standard
self-duality condition of Yang-Mills instantons in 8 dimensions. As a matter
of fact, Eq.~(\ref{gau_2_reduced}) is the defining relation of the
``octonionic instanton'' constructed in Ref.~\cite{Fubini:1985jm} and which
was used as source for the ``octonionic superstring soliton'' solution of the
Heterotic Superstring of Ref.~\cite{Harvey:1990eg}. Since this equation is
just a necessary condition to have at least one supersymmetry, we notice that
all supersymmetric solutions of the Heterotic Superstring effective action
must satisfy it. In particular, the gauge field of the ``gauge 5-brane''
solution of Ref.~\cite{Strominger:1990et} (a SU$(2)$ BPST instanton
\cite{Belavin:1975fg}) must satisfy it and, indeed, the self-duality condition
on the gauge field strength of the BPST instanton as just the result of
imposing the condition Eq.~(\ref{gau_2_reduced}) on a gauge field that lives
on a 4-dimensional subspace. For gauge fields that live in subspaces of
dimensions larger than 4 and smaller than 8, Eq.~(\ref{gau_2_reduced}) defines
Yang-Mills instantons of gauge groups related to the holonomy of the Killing
spinors. An intermediate example between the octonionic (Spin$(7)$) one and
the BPST one is provided by the the G$_{2}$ instanton and its associated
heterotic string solution \cite{Gunaydin:1995ku}.

\subsection{Summary of the necessary conditions for unbroken supersymmetry}
\label{sec-d10summary}

\begin{enumerate}
\item The metric has to admit a null Killing vector $\ell^{a}$. If $v$
  is the null coordinate adapted to this isometry, this means that the
  metric can be written in the form Eq.~(\ref{eq:metric}) which we
  rewrite here (as we will do with other formulae) for the sake of
  convenience

\begin{equation}
ds^{2}
=
2f(du+\beta)(dv+Kdu+\omega)
-h_{\underline{m}\underline{n}}dx^{m} dx^{n}\, ,
\hspace{1cm}
\beta = \beta_{\underline{m}}dx^{m}\, ,
\hspace{1cm}
\omega = \omega_{\underline{m}}dx^{m}\, .
\end{equation}

\noindent
All  objects in the metric are $v$-independent.

\item There exists a $v$-independent 4-form $\Omega$ satisfying the
  properties Eqs.~(\ref{eq:O4wedgeO4})-(\ref{eq:O2bis}) and which satisfies

  \begin{subequations}
    \begin{align}
      \label{eq:d++Omega}
      \nabla^{(+)}_{+}\Omega_{m_{1}\cdots m_{4}}
      & =
        0\, ,
      \\
      \nonumber \\
            \label{eq:d+nOmega}
      \nabla^{(+)}_{n}\Omega_{m_{1}\cdots m_{4}}
      & =
        0\, .
    \end{align}
  \end{subequations}

\item The following relations between certain components of the matter
  fields must be satisfied:

  \begin{subequations}
    \begin{align}
      \label{eq:d-phi}
      \partial_{-}\phi
      & =
        0\, ,
      \\
      \nonumber \\
      \label{gau_1-bis}
      F^{A}{}_{a-}
      & =
        0\, ,
      \\
      \nonumber \\
      \label{gau_2_reduced-bis}
      F^{A}{}_{mn}
      & =
        F^{A(+)}{}_{mn}\, ,
      \\
      \nonumber \\
      \label{eq:H-mn}
      H_{-mn}
      & =
        H^{(+)}_{-mn}\, ,
      \\
      \nonumber \\
      \label{eq:H-mnp-bis}
        H^{(-)}_{mnp}
      & =
        \tfrac{1}{7}(2\partial_{q}\phi -H_{+-q})\Omega^{q}{}_{mnp}\, ,
    \end{align}
  \end{subequations}
  
\item The torsionful spin connection $\Omega^{(+)}$ satisfies the
  following conditions:

  \begin{subequations}
    \begin{align}
      \label{eq:Omega+[ab]-}
      \Omega^{(+)}{}_{[ab]-}
      & =
        0\, ,
            \\
      \nonumber \\
      \label{eq:Omega+am-}
      \Omega^{(+)}{}_{am-}
      & =
        0\, ,
            \\
      \nonumber \\
      \label{eq:Omega+-mn}
      \Omega^{(+)}{}_{-mn}
      & =
         H^{(+)}_{-mn}\, .
    \end{align}
  \end{subequations}

  These conditions relate certain components of the Levi-Civita spin
  connection (and, hence, some of the objects that occur in the
  metric) to certain components of the Kalb-Ramond 3-form.

\end{enumerate}

\subsection{Sufficiency}
\label{sec-sufficiency}

Let us now check that the necessary conditions for
having the minimal amount of unbroken supersymmetry previously
identified are also sufficient.

\subsubsection{Gaugino KSE}
\label{sec-sufficiencygauginoKSE}

Let us start with the gaugino KSE Eq.~(\ref{KSE_gaugino}). The
necessary conditions that the gauge field strength has to satisfy are
Eq.~(\ref{gau_1-bis}) (\ref{gau_2_reduced-bis}). Then,

\begin{equation}
  F^{A}{}_{ab}\Gamma^{ab}\epsilon
  =
  F^{A(+)}{}_{mn}\Gamma^{mn}\Pi^{(-)}\epsilon
  +2F^{A}{}_{m+}\Gamma^{m}\Gamma^{+}\epsilon\, .
\end{equation}

\noindent
where we have used the property Eq.~(\ref{eq:F+property}) and the
spinor projector Eq.~(\ref{eq:Pi-spinorprojector}). This equation is
solved by demanding\footnote{The constraint Eq.~(\ref{eq:projector+})
  is associated to the projector
  \begin{equation}
  \tfrac{1}{2}\Gamma^{-}\Gamma^{+}  = \tfrac{1}{2}(1+\Gamma^{-+})\, .
  \end{equation}
}

\begin{subequations}
  \begin{align}
  \label{eq:projector+}
    \Gamma^{+}\epsilon
    & =
      0\, ,
    \\
    \nonumber \\
    \label{eq:Pi-epsilon}
    \Pi^{-}\epsilon
    & =
      0\, .
  \end{align}
\end{subequations}

\noindent
Observe that, when $F^{A}{}_{m+}=0$ the first condition seems to be
unnecessary. However, $\Pi^{-}$ is only idempotent when that condition
is satisfied.

\subsubsection{Dilatino KSE}
\label{sec-sufficiencydilatinoKSE}

Using Eq.~(\ref{eq:d-phi}) and the spinor projector
Eq.~(\ref{eq:projector+}), this equation reduces to

\begin{equation}
\label{eq:dilatonKSEsolving-1}
  \left\{  \left(\partial_{m}\phi -\tfrac{1}{2}H_{+-m}\right)\Gamma^{m}
  -\tfrac{1}{12}H_{mnp}\Gamma^{mnp} -\tfrac{1}{4}H_{-mn}\Gamma^{-}\Gamma^{mn}
\right\}\epsilon
=
0\, .
\end{equation}

Now we do two things:

\begin{enumerate}
\item First use Eq.~(\ref{eq:H-mnp-bis}) into Eq.~(\ref{eq:Hproperty})
  to eliminate $\Pi^{+}H_{mnp}$ from the latter, solve the resulting
  equation for $H_{mnp}$ and substitute the result in the above equation.
  
\item Use Eq.~(\ref{eq:H-mn}) and then Eq.~(\ref{eq:F+property}).
  
\end{enumerate}

Eq.~(\ref{eq:dilatonKSEsolving-1}) takes the form

\begin{equation}
  \label{eq:dilatonKSEsolving-2}
  \left\{\tfrac{7}{8}  \left(\partial_{m}\phi -\tfrac{1}{2}H_{+-m}\right)
    \left(\Gamma^{m}-\tfrac{1}{42}\Omega^{m}{}_{npq}\Gamma^{npq}\right)
    -\tfrac{1}{12}\left(H_{mnp}\Gamma^{mnp}
    +3H_{-mn}\Gamma^{-}\Gamma^{mn}\right)\Pi^{(-)}
\right\}\epsilon
=
0\, .
\end{equation}

\noindent
Then, Eq.~(\ref{eq:GammamPi-}) allows us to rewrite the whole equation
as

\begin{equation}
  \label{eq:dilatonKSEsolving-3}
  \left\{\Gamma^{m}\partial_{m}\phi  -\tfrac{1}{12}\left(H_{mnp}\Gamma^{mnp}
    +3H_{-mn}\Gamma^{-}\Gamma^{mn}+6H_{+-m}\Gamma^{m}\right)
\right\}\Pi^{(-)}\epsilon
=
0\, ,
\end{equation}

\noindent
which is solved by demanding Eq.~(\ref{eq:Pi-epsilon}).

\subsubsection{Gravitino  KSE}
\label{sec-sufficiencygravitinoKSE}

The projection Eq.~(\ref{eq:projector+}) and the supersymmetry conditions
Eqs.~(\ref{eq:Omega+[ab]-}) and (\ref{eq:Omega+am-}) bring the gravitino KSE
to the form\footnote{Observe that, in the basis we are using, we also have
  $\Omega^{(+)}{}_{+\, +-}=\omega_{++-}=0$. Then, combining this result with
  Eqs.~(\ref{eq:Omega+[ab]-}) and (\ref{eq:Omega+am-}), we get
  $\Omega^{(+)}{}_{a\, +-}=0$.}

\begin{equation}
  \label{eq:gravitinoKSEsolving-1}
  \left(\partial_{a}-\tfrac{1}{4}\Omega^{(+)}{}_{a\, mn}\Gamma^{mn}\right)
  \epsilon=0\, .
\end{equation}

Due to the condition Eq.~(\ref{eq:Omega+-mn}) and of the property
Eq.~(\ref{eq:F+property}) and of the projection Eq.~(\ref{eq:Pi-epsilon}), the
$a=-$ component is solved by $v$-independent spinors

  \begin{equation}
  \left(\partial_{-}
    -\tfrac{1}{4}  H^{(+)}{}_{-mn}\Gamma^{mn}
  \right)\epsilon
  =
  \partial_{-}\epsilon
  =
  0\,.
\end{equation}

The $a=+,m$ components are also guaranteed to be satisfied because of the
supersymmetry conditions Eqs.~(\ref{eq:d++Omega}) and
(\ref{eq:d+nOmega}). Observe that these two equations lead to

\begin{equation}
  \Omega^{(+)}{}_{a\, mn}
  =
  \left(\Pi^{(+)}\Omega^{(+)}{}_{a}\right)_{mn}
  +\tfrac{1}{96}\partial_{b}\Omega_{mp_{1}p_{2}p_{3}} \Omega_{n}{}^{p_{1}p_{2}p_{3}}\, .
\end{equation}

If we use a frame in which the components of the 4-form are constant, then 
Eq.~(\ref{eq:gravitinoKSEsolving-1}) takes the form

\begin{equation}
  \label{eq:gravitinoKSEsolving-2}
  \left(\partial_{a}
    -\tfrac{1}{4}  \left(\Pi^{(+)}\Omega^{(+)}{}_{a}\right)_{mn}\Gamma^{mn}
  \right)\epsilon
  =
  \partial_{a}\epsilon
  =
  0\, , 
\end{equation}

\noindent
by virtue of the property Eq.~(\ref{eq:F+property}) and of the
projection Eq.~(\ref{eq:Pi-epsilon}). Then, in that frame, the Killing
spinors are just constant spinors satisfying the two conditions
Eqs.~(\ref{eq:projector+}) and (\ref{eq:Pi-epsilon}).

\section{Supersymmetric solutions}
\label{sec-d10hetsol}

In this section we are going to study under which conditions, the
supersymmetric field configurations that we have identified in the previous
section are also solutions of the equations of motion of the theory. We start
by reviewing the equations of motion that follow from the action
Eq.~(\ref{heterotic}) at first order in $\alpha'$ and finding the relations
with the simplified equations which are usually solved. Of course, nothing but
the sheer difficulty prevents us from deriving higher-order equations of
motion from Eq.~(\ref{heterotic}) because it is not known if the
simplifications that occur at first order in $\alpha'$ have an analogue at
higher orders.

\subsection{Equations of motion}
\label{sec-EOM}

The equations of motion that follow from the action Eq.~(\ref{heterotic}) are
very complicated. If we stay at first order in $\alpha'$, though, there are
important simplifications. Following Ref.~\cite{Bergshoeff:1992cw}, we can
separate the variations with respect to each field in the action (except for
the dilaton and Yang-Mills fields) into those corresponding to occurrences via
${\Omega}_{(-)}{}^{{a}}{}_{{b}}$, that we will call
\textit{implicit},\footnote{The dilaton does not occur in the torsionful spin
  connection, neither do the Yang-Mills fields to the order in $\alpha'$ we
  are considering.} and the rest, that we will call \textit{explicit}, as
follows:\footnote{The complete equations of motion of all the fields and
  $\delta S/\delta {\Omega}_{(-)\, \mu}{}^{{a}}{}_{{b}}$ can be found in
  Appendix~\ref{app-EOM} to first order in $\alpha'$. The proof of the lemma
  of Ref.~\cite{Bergshoeff:1992cw} mentioned below can also be found there.}

\begin{eqnarray}
\delta S 
& = &  
\frac{\delta S}{\delta e^{a}{}_{\mu}}\delta e^{a}{}_{\mu}
+\frac{\delta S}{\delta B_{\mu\nu}}\delta B_{\mu\nu}
+\frac{\delta S}{\delta A^{A}{}_{\mu}}\delta A^{A}{}_{\mu}
+\frac{\delta S}{\delta \phi} \delta \phi
\nonumber \\
& & \nonumber \\
& = & 
\left.\frac{\delta S}{\delta g_{\mu\nu}}\right|_{\rm exp.}\delta g_{\mu\nu}
+\left.\frac{\delta S}{\delta B_{\mu\nu}}\right|_{\rm exp.}\delta B_{\mu\nu}
+\frac{\delta S}{\delta A^{A}{}_{\mu}}\delta A^{A}{}_{\mu}
+\frac{\delta S}{\delta \phi} \delta \phi
\nonumber \\
& & \nonumber \\
& &
+\frac{\delta S}{ \delta {\Omega}_{(-)\, \mu}{}^{{a}}{}_{{b}}}
\left(
\frac{\delta {\Omega}_{(-)\, \mu}{}^{{a}}{}_{{b}}}{\delta e^{c}{}_{\rho}}\delta e^{c}{}_{\rho}
+\frac{\delta {\Omega}_{(-)\, \mu}{}^{{a}}{}_{{b}}}{\delta B_{\nu\rho}} \delta B_{\nu\rho}
\right)\, .
\end{eqnarray}

A lemma proven in Ref.~\cite{Bergshoeff:1989de} states that
$\delta S/\delta {\Omega}_{(-)\, \mu}{}^{{a}}{}_{{b}}$ is proportional to
$\alpha'$ multiplied by combinations of zeroth-order equations of motion of
the fields $e^{a}{}_{\mu},B_{\mu\nu}$ and $\phi$ plus terms of higher orders
in $\alpha'$.

This lemma has important consequences: if we consider field configurations
which solve the zeroth-order equations of motion up to terms of order
$\alpha'$ or higher, then
$\delta S/\delta {\Omega}_{(-)\, \mu}{}^{{a}}{}_{{b}}=X^{\mu}{}_{a}{}^{b}$
will automatically vanish up to terms of second order in $\alpha'$ or higher
and can be ignored to the order at which we are working. Then all the terms
involving $X^{\mu}{}_{ab}$ in the complete equations of motion
Eqs.~(\ref{eq:completedilatoneom})-(\ref{eq:completevielbeineom}) can be
ignored.

Further simplifications are possible combining the equations of motion:

\begin{eqnarray}
\label{eq:eq1}
   - \frac{e^{2\phi}}{2 e}
    \left[\frac{\delta S}{\delta e^{a}{}_{\mu}}
    +\tfrac{1}{2}e_{a}{}^{\mu}\frac{\delta S}{\delta \phi}\right]_{\rm exp.}
& = &
  R^{\mu}{}_{a} -2\nabla^{\mu}\partial_{a}\phi
+\tfrac{1}{4}{H}^{\mu\rho\sigma}{H}_{a\rho\sigma}
-T^{(2)\, \mu}{}_{a}\, ,
\\
& & \nonumber \\
\label{eq:eq2}
   \frac{e^{2\phi}}{4 e}
    \left[e^{a}{}_{\mu}\frac{\delta S}{\delta e^{a}{}_{\mu}}
    +\frac{d-2}{2}\frac{\delta S}{\delta \phi}\right]_{\rm exp.}
& = &
(\partial \phi)^{2} -\tfrac{1}{2}\nabla^{2}\phi
-\tfrac{1}{4\cdot 3!}{H}^{2}
+\tfrac{1}{8}T^{(0)}\, ,
\\
& & \nonumber \\
  \label{eq:eq3}
   -\frac{2}{e}\left.\frac{\delta S}{\delta B_{\nu\rho}}\right|_{\rm exp.}
& = &
  \nabla_{\mu}\left(e^{-2\phi}H^{\mu\nu\rho}\right)\, ,
\\
& & \nonumber \\
  \label{eq:eq4}
  \frac{2}{e}\left[\frac{\delta S}{\delta A^{A}{}_{\nu}}
  +A^{A}{}_{\mu}\left.\frac{\delta S}{\delta B_{\mu\nu}}\right|_{\rm exp.}\right]
  & = &
\alpha' e^{2\phi}\nabla_{(+)\, \mu}\left(e^{-2\phi}F^{A\, \mu\nu}\right)\, .
\end{eqnarray}

These are the equations of motion that are usually solved in the
literature. Observe that, although we have arrive at them by assuming that the
zeroth-order equations of motion are satisfied, since these can be obtained by
setting $\alpha'=0$ in Eqs.~(\ref{eq:eq1})-(\ref{eq:eq4}), if we find a
solution of Eqs.~(\ref{eq:eq1})-(\ref{eq:eq4}), then it is automatically a
solution of the zeroth-order equations of motion up to terms of first order in
$\alpha'$, the lemma can be applied, the complete equations of motion reduce
to Eqs.~(\ref{eq:eq1})-(\ref{eq:eq4}) and the solution is a solution of the
complete equations of motion.

The non-trivial relation between the complete equations of motion that one
obtains from the action
Eqs.~(\ref{eq:completedilatoneom})-(\ref{eq:completevielbeineom}) and the
standard equations of motion considered in the literature
Eqs.~(\ref{eq:eq1})-(\ref{eq:eq4}) that we have just explained has to be kept
in mind when using the Killing Spinor Identities (KSIs)
\cite{Kallosh:1993wx,Bellorin:2005hy} because, in principle, they involve the
complete ones
Eqs.~(\ref{eq:completedilatoneom})-(\ref{eq:completevielbeineom}) (see
Section~\ref{sec-KSIs}).

If a solution is given in terms of the 3-form field strength, we also need to
solve the Bianchi identity

\begin{equation}
\label{eq:BianchiH}
d{H}  
-
T^{(4)}
=
0\, ,
\end{equation}

\noindent
as well. 

This identity can be rewritten as the equation of motion of the 6-form
$\tilde{B}_{\mu_{1}\cdots\mu_{6}}$ dual to the Kalb-Ramond 2-form:

\begin{equation}
  \nabla_{\nu}\left(\tilde{H}^{\nu\mu_{1}\cdots\mu_{6}}\right)
  -\star T^{(4)\, \mu_{1}\cdots\mu_{6}}
  =
  0\, ,
\end{equation}

\noindent
where $\tilde{H}=d\tilde{B}=e^{-2\phi}\star H$ is its field strength. Then, we
can introduce it in the Killing Spinor Identities (see Section~\ref{sec-KSIs})
which only involve equations of motion if we use the dual formulation of the
10-dimensional supergravity in
Refs.~\cite{Bergshoeff:1990ax,Bergshoeff:1990hh} if we take into account that
the above equation corresponds to

\begin{equation}
  \frac{6!}{\sqrt{|g|}}\frac{\delta S}{\delta \tilde{B}_{\mu_{1}\cdots\mu_{6}}}
  =
  0\, .
\end{equation}

Apart from the Kalb-Ramond 2-form, there are no other fields in the action
which only occur through their field strengths. Thus, we cannot formulate the
equations of motion only in terms of those field strengths and imposing a
Bianchi identity on them is utterly unnecessary. Observe that, for instance,
in order to write the Bianchi identity for the Yang-Mills field strength (or
the Riemann curvature tensor) it is necessary to know the gauge field (the
connection). Therefore, we will not need any more equations.

\subsection{Killing Spinor Identities}
\label{sec-KSIs}

The equations of motion of theories with local symmetries are related
off-shell by the so-called Noether (or gauge) identities. In a supergravity
theory the Noether identities relate the bosonic and fermionic equations of
motion.\footnote{The invariance of the action of $N=1,d=4$ supergravity was
  first proven analytically in Ref.~\cite{Deser:1976eh} precisely by checking
  that the supersymmetric Noether identity was satisfied off-shell.} These
identities are valid for any field configurations but, if we restrict
ourselves to purely bosonic field configurations admitting Killing spinors,
such as those we have characterized in the previous section, it can be shown
that the equations of motion of the bosonic fields are related by the
so-called \textit{Killing Spinor Identities}, first derived in
Ref.~\cite{Kallosh:1993wx}.

As shown in Ref.~\cite{Bellorin:2005hy}, the KSIs are essentially equivalent
to the relations obtained from the integrability conditions of the Killing
spinor equations, but, in general, they are much easier to derive because,
usually, only algebraic operations are required. These relations between the
bosonic equations of motion of supersymmetric configurations can be used to
reduce the number of independent equations that need to be checked in order to
prove that a given supersymmetric field configuration is also a solutions of
all equations of motion. Our goal in this section is to find the KSIs and
determine the independent equations of motion that need to be checked in the
case of the Heterotic Superstring effective action.

An important point to be stressed is that the generic form of the KSIs only
depends on the supersymmetry transformation laws of the bosonic fields. The
equations of motion of a supergravity theory change when we gauge it, deform
it with mass terms or, as it is the case here, when we add the $\alpha'$
corrections, but, in many cases, the relations that hold between them when
they are evaluated on supersymmetric configurations, the KSIs, do not, because
they only depend on the supersymmetry transformation rules of the bosonic
fields, which do not change.\footnote{This observation was used in
  Ref.~\cite{Meessen:2007ef} to prove the exactness of the maximally
  supersymmetric solutions of 5-dimensional supergravity when higher-order
  corrections are included.}

In the present case, however, the supersymmetry transformation rules of the
bosons do get $\alpha'$ corrections and this generic property will not hold
true. In particular, as explained in the Introduction, the Kalb-Ramond 2-form
supersymmetry transformation rules acquire two new terms of first order in
$\alpha'$ (see Eq.~(\ref{eq:susyruleKR})) associated to the Nicolai-Townsend
(``anomalous'') gauge transformations that arise when the Lorentz- and
Yang-Mills-Chern-Simons terms are included in the field strength. At higher
orders, new terms have to be taken into account and, although it has been
suggested that they may be absorbed in redefinitions of the fields, this has
not been explored systematically and we will restrict our analysis to the
first order in $\alpha'$ at which the only modification of the bosons'
supersymmetry transformation rules is the one we have just discussed.

A disadvantage of this approach that the proof of the KSIs in
Ref.~\cite{Kallosh:1993wx} assumes the existence of the potentials in the
field strengths (the equations of motion are the first variations of the
action with respect to them) or, equivalently, that the Bianchi identities are
satisfied and, sometimes, we would like not to assume this and solve a
different set of equations. The Bianchi identities appear explicitly in the
integrability conditions, but it is usually very hard to compute them.

There is, however, a simple way to make the Bianchi identities appear in the
KSIs: we just have to view them as the equations of motion of the dual
potentials (as long as their supersymmetry transformation laws are known). In
the case at hands, this means that, if we want to find KSIs including the
Bianchi identity of the Kalb-Ramond 3-form field strength, we must view it as
the equation of motion of the dual 6-form potential
$\tilde{B}_{\mu_{1}\cdots\mu_{6}}$ and use the supersymmetry transformation
law of this field, given in Refs.~\cite{Bergshoeff:1990ax,Bergshoeff:1990hh}
and which we have rewritten in our conventions in
Eq.~(\ref{eq:susytrans6-form}).  Observe that, if denote by
$\mathcal{E}_{\tilde{B}}{}^{\mu_{1}\cdots\mu_{6}}$ the equation of motion of
the dual 6-form $\tilde{B}_{\mu_{1}\cdots\mu_{6}}$ (see below), and we denote
by $\mathcal{B}_{H\, \mu_{1}\cdots \mu_{4}}$ the Bianchi identity of the
Kalb-Ramond 3-form,\footnote{Up to global factors, the components of the
  Bianchi identity Eq.~(\ref{eq:BianchiH}) are
  \begin{equation}
    \mathcal{B}_{H\, \mu_{1}\cdots \mu_{4}}
    =
  4\partial_{[\mu_{1}}H_{\mu_{2}\mu_{3}\mu_{4}]}
  -T^{(4)}{}_{\mu_{1}\cdots \mu_{4}}\, .
\end{equation}
}

\noindent
they are each other's Hodge dual:

\begin{equation}
  \label{eq:6versus4}
  \mathcal{E}_{\tilde{B}}{}^{\mu_{1}\cdots\mu_{6}}
  =\frac{1}{4!\sqrt{|g|}}\epsilon^{\mu_{1}\cdots\mu_{6}\nu_{1}\cdots\nu_{4}}
  \mathcal{B}_{H\, \nu_{1}\cdots \nu_{4}}\, .
\end{equation}

Taking into account this last point, using the definitions

\begin{equation}
  \begin{aligned}
    \mathcal{E}_{e}{}^{\mu}{}_{a}
    & \equiv &
            \left.\frac{\delta S}{\delta e^{a}{}_{\mu}}\right|_{\rm exp}\, ,
               \hspace{1cm}
               &
                 \mathcal{E}_{B}{}^{\mu\nu}
    & \equiv &
    \left.\frac{\delta S}{\delta B_{\mu\nu}}\right|_{\rm exp}\, ,
    \hspace{1cm}
               &
                 \mathcal{E}_{\tilde{B}}{}^{\mu_{1}\cdots\mu_{6}}
    & \equiv &
              \frac{\delta S}{\delta
               \tilde{B}_{\mu_{1}\cdots\mu_{5}}}\, , 
    \\
    & & & & & & & & \\
    \mathcal{E}_{\phi}
    & \equiv  &
                \frac{\delta S}{\delta \phi}\, ,  \hspace{2cm}
                &
                  \mathcal{E}_{A}{}^{\mu}
    & \equiv  &
                 \frac{\delta S}{\delta A^{A}{}_{\mu}}\, ,
                 &
                 X^{\mu}{}_{ab}
                 & \equiv & \frac{1}{\sqrt{|e|}}
                 \frac{\delta S}{\delta \Omega_{(-)\, \mu}{}^{ab}}\, ,
\\
  \end{aligned}
\end{equation}

\noindent
and splitting the variations with respect to the fields into variations with
respect to explicit occurrences and variations with respect to implicit
occurrences via the torsionful spin connection,\footnote{This step is
  fundamental to recover the standard equations of motion
  Eqs.~(\ref{eq:eq1})-(\ref{eq:eq4}).} the general recipe in
Ref.~\cite{Kallosh:1993wx} takes the form

\begin{subequations}
  \begin{align}
    \mathcal{E}_{e\, a}{}^{\nu}
        \frac{\delta(\delta_{\epsilon} e^{a}{}_{\nu})}{\delta \psi_{\mu}}
      +\mathcal{E}_{B}{}^{\nu\rho}
  \frac{\delta(\delta_{\epsilon} B_{\nu\rho})}{\delta \psi_{\mu}} 
    +\mathcal{E}_{\tilde{B}}{}^{\mu_{1}\cdots\mu_{6}}
    \frac{\delta ( \delta_{\epsilon} \tilde{B}_{\mu_{1}\cdots\mu_{6}})}{\delta\psi_{\mu}}
+\sqrt{|g|}X^{\nu}{}_{ab}\frac{\delta(\delta_{\epsilon} \Omega_{(-)\, \nu}{}^{ab})}{\delta \psi_{\mu}}
    & =
0\, ,
    \\
    \nonumber \\
    \mathcal{E}_{\phi}
    \frac{\delta(\delta_{\epsilon} \phi)}{\delta \lambda}
+\mathcal{E}_{\tilde{B}}{}^{\mu_{1}\cdots\mu_{6}}
    \frac{\delta ( \delta_{\epsilon} \tilde{B}_{\mu_{1}\cdots\mu_{6}})}{\delta\lambda}    
& =
0\, ,
    \\
    \nonumber \\
  \mathcal{E}_{C}{}^{\mu}
    \frac{\delta(\delta_{\epsilon} A^{C}{}_{\mu})}{\delta \chi^{A}}
    +\mathcal{E}_{B}{}^{\mu\nu}
    \frac{\delta(\delta_{\epsilon} B_{\mu\nu})}{\delta \chi^{A}}
& =
0\, .    
  \end{align}
\end{subequations}

Using the supersymmetry transformations of the bosons in
Eqs.~(\ref{eq:susyrule1})-(\ref{eq:susyrule5}), we find that the KSIs of
the theory at hands are the spinorial equations

\begin{subequations}
  \begin{align}
  \label{KSI_e_B-1}
    \bar{\epsilon}
    \left\{\Gamma^{a}  \mathcal{E}_{e\, a}{}^{\mu}
    +2\Gamma_{\alpha}\mathcal{E}_{B}{}^{\alpha\mu}
    +6\Gamma_{\alpha_{1}\cdots\alpha_{5}}
    e^{-2\phi}\mathcal{E}_{\tilde{B}}{}^{\alpha_{1}\cdots\alpha_{5}\mu}
    \right\}
    -\mathcal{D}_{(+)\, \lambda}\left(\bar{\epsilon}\Gamma_{a}Y^{a\lambda\mu}\right)
    & =
      0\, , 
    \\
    \nonumber \\
    \label{KSI_phi}
    \bar{\epsilon}
    \left\{
    \mathcal{E}_{\phi}
    -2\Gamma_{\alpha_{1}\cdots\alpha_{6}}
    e^{-2\phi}\mathcal{E}_{\tilde{B}}{}^{\alpha_{1}\cdots\alpha_{6}}
    \right\}
& =
0\, ,
    \\
    \nonumber \\
    \label{KSI_A-1}
    \bar{\epsilon}\Gamma_{\alpha}\left(
    \mathcal{E}_{A}{}^{\alpha}
    +\frac{\alpha'}{2}A_{A\, \mu}\mathcal{E}_{B}{}^{\mu\alpha}
    \right)
& =
0\, ,    
  \end{align}
\end{subequations}

\noindent
where we have defined the combination

\begin{equation}
  Y^{\mu}{}_{ab}
  \equiv
  \sqrt{|g|}X^{\mu}{}_{ab}
  -\frac{\alpha'}{2}\mathcal{E}_{B}{}^{\rho \mu}\Omega_{(-)\, \rho\, ab}\, .
\end{equation}

\noindent
Using Eq.~(\ref{eq:Xcombination}), 

\begin{equation}
  Y^{\nu}{}_{ab}
  \equiv
    -\frac{\alpha'}{2}e^{-2\phi}
      \nabla^{(0)}_{(+)\, [a|}
      \left[
        e^{2\phi}
  \left(\mathcal{E}^{(0)}_{e\, |b]}{}^{\nu}+2\mathcal{E}^{(0)}_{B\, |b]}{}^{\nu}
    +\tfrac{1}{2}e_{|b]}{}^{\nu}\mathcal{E}^{(0)}_{\phi}
  \right)
        \right]\, ,
\end{equation}

\noindent
and, taking into account that the Killing spinor $\epsilon$ satisfies
$\nabla_{(+)}^{(0)}{}_{\lambda}\epsilon=\mathcal{O}(\alpha')$, we find that

\begin{equation}
    \bar{\epsilon}\Gamma_{a}Y^{a\lambda\mu}
    =  -\frac{\alpha'}{2} 
      e^{-2\phi} \nabla^{(0)}_{(+)}{}^{[\lambda|} \left[ e^{2\phi}
        \bar{\epsilon}\Gamma_{a}\left(\mathcal{E}_{e}^{(0)\, |\mu]
            a}+2\mathcal{E}_{B}^{(0)\, |\mu] a} +\tfrac{1}{2}e^{a\,
            |\mu]}\mathcal{E}^{(0)}_{\phi} \right) \right]\, ,
\end{equation}

\noindent
up to terms of order $\mathcal{O}(\alpha'^{2})$. Now we can use the KSI
Eq.~(\ref{KSI_e_B-1}) at zeroth order in $\alpha'$ to show that

\begin{equation}
  e^{2\phi}\bar{\epsilon}\Gamma_{a}\left(\mathcal{E}_{e}^{(0)\, a\mu}
    +2\mathcal{E}_{B}^{(0)\, a \mu} +\tfrac{1}{2}e^{a \mu}\mathcal{E}^{(0)}_{\phi} \right)
        =
        \bar{\epsilon}\Gamma_{\alpha_{1}\cdots \alpha_{6}}{}^{\mu}
        \mathcal{E}^{(0)}_{\tilde{B}}{}^{\alpha_{1}\cdots\alpha_{6}}\, ,
\end{equation}

\noindent
so that

\begin{equation}
    \bar{\epsilon}\Gamma_{a}Y^{a\lambda\mu}
    =  -\frac{\alpha'}{2} 
    e^{-2\phi} \nabla^{(0)}_{(+)}{}^{[\lambda}
    \left(\bar{\epsilon}\Gamma_{\alpha_{1}\cdots \alpha_{6}}{}^{\mu]}
        \mathcal{E}^{(0)}_{\tilde{B}}{}^{\alpha_{1}\cdots\alpha_{6}}\right)\, .
\end{equation}

The upshot is that, to $\mathcal{O}(\alpha'^{2})$, the KSE
Eq.~(\ref{KSI_e_B-1}), can be replaced by

\begin{equation}
  \label{KSI_e_B}
    \bar{\epsilon}
    \left\{\Gamma^{a}  \mathcal{E}_{e\, a}{}^{\mu}
    +2\Gamma_{\alpha}\mathcal{E}_{B}{}^{\alpha\mu}
    +6\Gamma_{\alpha_{1}\cdots\alpha_{5}}
    e^{-2\phi}\mathcal{E}_{\tilde{B}}{}^{\alpha_{1}\cdots\alpha_{5}\mu}
  \right\}
  +\bar{\epsilon}\left(\Gamma^{(7)}C\right)^{\mu}
  =
  0\, ,
\end{equation}

\noindent
where $C$ is a complicated tensor that contains $\mathcal{E}^{(0)}_{\tilde{B}}$ and
$\Gamma^{(7)}$ is the antisymmetrized product of 7 gamma matrices.

Furthermore, observe that the combination of equations of motion that occurs
in Eq.~(\ref{KSI_A-1})

\begin{equation}
\mathcal{E}_{A}{}^{\alpha}
+\frac{\alpha'}{2}A_{A\, \mu}\mathcal{E}_{B}{}^{\mu\alpha}
=
\frac{\alpha'}{2}\sqrt{|g|} \nabla_{(+)\, \nu}\left(e^{-2\phi}F^{A\,
    \nu\mu}\right)
\equiv
\hat{\mathcal{E}}_{A}{}^{\alpha}\, ,
\end{equation}

\noindent
and it is convenient to rewrite the KSI in the form

\begin{equation}
\label{KSI_A}
    \bar{\epsilon}\Gamma_{\alpha}
    \hat{\mathcal{E}}_{A}{}^{\alpha}
    =
    0\, .    
\end{equation}

Eqs.~(\ref{KSI_phi}), (\ref{KSI_e_B}) and (\ref{KSI_A}) should be identical to
the relations between the bosonic equations of motion obtained from the KSEs'
integrability conditions.  These have been recently presented in
Ref.~\cite{Gran:2018ijr} (Eqs.~(205)) and should be compared to
Eqs.~(\ref{KSI_phi}), (\ref{KSI_e_B}) and (\ref{KSI_A}). Apart from some minor
differences (the equation of motion of the dual 6-form is replaced by the
Bianchi identity of the 3-form and the Einstein and Maxwell equations used in
Ref.~\cite{Gran:2018ijr} are the combinations of our equations of motion
described in Eqs.~(\ref{eq:eq1})-(\ref{eq:eq4})) the authors of
Ref.~\cite{Gran:2018ijr} have used an expansion of the equations of motion and
Bianchi identities in powers of $\alpha'$ that, for instance, does not take
into account the definition of the Kalb-Ramond field strength $H$ in terms of
potentials Eq.~(\ref{eq:H1def}) and the corresponding expansion in
$\alpha'$. Instead, they just assume the existence of field strengths which
are not expanded in powers of $\alpha'$, satisfying certain
(anomaly-corrected) Bianchi identities . Using the explicit form of the
Bianchi identity for $H$ it is possible to rewrite their Eqs.~(205) in exactly
the same form we have obtained.

\subsection{Tensorial KSIs}
  
Eqs.~(\ref{KSI_phi}), (\ref{KSI_e_B}) and (\ref{KSI_A}) are the off-shell
relations between the bosonic equations of motion we were after but, in order
to make use of them, we must transform them into tensorial equations. Let us
start with the simplest of them, Eq.~(\ref{KSI_A}): if we hit it with
$\epsilon$ and $\Gamma^{\mu_{1}\mu_{2}\mu_{3}\mu_{4}} \epsilon$ from the right
we obtain, respectively,

\begin{subequations}
  \begin{align}
    \label{eq:KSI1}
  \ell_{\mu} \hat{\mathcal{E}}_{A}{}^{\mu}
  & =
    0\, ,
  \\
  & \nonumber \\
  W_{\mu_{1}\mu_{2}\mu_{3}\mu_{4} \mu} \hat{\mathcal{E}}_{A}{}^{\mu}
  & =
    0\, .
\end{align}
\end{subequations}

\noindent
Using Eq.~(\ref{eq:WlO}) and the first equation in the second, and contracting
it with the null vector $n$, it takes the simpler form

\begin{equation}
  \Omega_{\mu_{1}\mu_{2}\mu_{3}\mu} \hat{\mathcal{E}}_{A}{}^{\mu}
 =
    0\, .
\end{equation}

\noindent
Contracting this equation with $\Omega^{\mu_{1}\mu_{2}\mu_{3} \nu}$ and using
Eq.~(\ref{eq:O3O3eta}), we get

\begin{equation}
  \label{eq:KSI2}
  \tilde{g}_{\mu\nu}\hat{\mathcal{E}}_{A}{}^{\nu} = 0 \, , 
\end{equation}

\noindent
where $\tilde{g}_{\mu\nu}=g_{\mu\nu}-2\ell_{(\mu}n_{\nu)}$ is the (curved
indices) induced metric in the 8-dimensional Euclidean transverse space
defined in Eq.~(\ref{eq:inducedmetric}). An equivalent way of writing this
equation is

\begin{equation}
  \label{eq:KSI2-bis}
\hat{\mathcal{E}}_{A}{}^{m} = 0 \, .   
\end{equation}

We conclude, that all  components of the Yang-Mills equations
$\hat{\mathcal{E}}_{A}{}^{\mu}$, except for $n_{\mu}\hat{\mathcal{E}}_{A}{}^{\mu}$, are
automatically satisfied by supersymmetric field configurations.

Hitting now Eq.~(\ref{KSI_phi}) with $\Gamma^{\mu}\epsilon$ from the right and
contracting the result with $n_{\mu}$, we arrive at

\begin{equation}
  \label{eq:KSI3}
  \mathcal{E}_{\phi}
  =
60\ell_{\mu}n_{\nu}\mathcal{E}_{\tilde{B}}{}^{\mu\nu}\, .
\end{equation}

\noindent
where we have defined 

\begin{equation}
  \mathcal{E}_{\tilde{B}}{}^{\mu\nu}
  \equiv
  \Omega_{\alpha_{1}\alpha_{2}\alpha_{3}\alpha_{4}}
e^{-2\phi}\mathcal{E}_{\tilde{B}}{}^{\alpha_{1}\alpha_{2}\alpha_{3}\alpha_{4}\mu\nu}\, ,  
\end{equation}

\noindent
because this combination appears very often and plays an interesting
r\^ole. Often, in the literature, the Bianchi identity is assumed to be solved
from the beginning. In that case, the dilaton equation is automatically solved
on supersymmetric field configurations, but the above KSI allows us to assume
that the dilaton equation is solved from the beginning, which would imply that
the component $\ell_{\mu}n_{\nu}\mathcal{E}_{\tilde{B}}{}^{\mu\nu}$ of the
Bianchi identity is automatically solved.

Observe that the relation Eq.~(\ref{eq:6versus4}) implies that 

\begin{equation}
  \ell_{\mu}n_{\nu}\mathcal{E}_{\tilde{B}}{}^{\mu\nu}
  =
  -e^{-2\phi}\Omega^{\alpha_{1}\cdots\alpha_{4}}
\mathcal{B}_{H\, \alpha_{1}\cdots\alpha_{4}}\, ,
\end{equation}

\noindent
so the dilaton equation is related to a single combination of the transverse
components of the Bianchi identity.

Next, let us consider Eq.~(\ref{KSI_e_B}). Hitting it with $\epsilon$ from the
right, and taking into account that the last term will not contribute, we get

\begin{equation}
  \label{eq:KSI4}
  \ell_{\mu} \left(\mathcal{E}_{e}{}^{\mu\nu}
  +2\hat{\mathcal{E}}_{B}{}^{\mu\nu}\right)
  =
  0\, ,
\end{equation}

\noindent
where we have defined 

\begin{equation}
  \hat{\mathcal{E}}_{B}{}^{\mu\nu}
  \equiv
  \mathcal{E}_{B}{}^{\mu\nu}+15\mathcal{E}_{\tilde{B}}{}^{\mu\nu}\, .
\end{equation}

\noindent
from which we get

\begin{equation}
  \label{eq:KSI5}
  \ell_{\mu}\ell_{\nu}\mathcal{E}_{e}{}^{\mu\nu}
  =
  0\, .
\end{equation}

Observe that, while the complete Einstein equation,
$e^{a\nu}\delta S/\delta e^{a}{}_{\mu}$, is not symmetric in the pair of
indices $\mu\nu$,
$\mathcal{E}_{e}{}^{\mu\nu} = e^{a\, \mu}\mathcal{E}_{e\, a}{}^{\nu}$, which
only contains the variation of the action with respect to explicit occurrences
of the Vielbein, is.\footnote{This fact follows from the Noether identities,
  see Eq.~(\ref{eq:noetheridentity5}).}

In what follows, in order to get rid of the complicated contributions of the
last term in Eq.~(\ref{KSI_e_B}) we will assume that the Bianchi identity is
satisfied at zeroth order in $\alpha'$,

\begin{equation}
  \mathcal{E}^{(0)}_{\tilde{B}}=0\, ,
  \hspace{1cm}
  \mathcal{E}_{\tilde{B}} = \mathcal{O}(\alpha')\, .
\end{equation}

Using this assumption and hitting Eq.~(\ref{KSI_e_B}) with
$\Gamma^{\rho}\Gamma^{\sigma}\epsilon$ from the right and using
Eq.~(\ref{eq:KSI4}) we get

\begin{equation}
\label{eq:KSIotra}
\left(\mathcal{E}_{e}^{[\rho|\mu}+2\hat{\mathcal{E}}_{B}^{[\rho|\mu}
\right)\ell^{\sigma]}
+120\ell_{\nu}e^{-2\phi}\mathcal{E}_{\tilde{B}}{}^{\alpha_{1}\alpha_{2}\alpha_{3}[\rho|\nu\mu}
\Omega_{\alpha_{1}\alpha_{2}\alpha_{3}}{}^{|\sigma]}
=
0\, .
\end{equation}

\noindent
If the antisymmetrized indices $\rho$ and $\sigma$ are transverse indices $m$
and $n$ the first term vanishes identically and we get 

\begin{equation}
\label{eq:KSIotra2}
\ell_{\nu}\mathcal{E}_{\tilde{B}}{}^{\alpha_{1}\alpha_{2}\alpha_{3}[m|\nu\mu}
\Omega_{\alpha_{1}\alpha_{2}\alpha_{3}}{}^{|n]}
=
0\, .
\end{equation}

Furthermore, hitting Eq.~(\ref{eq:KSIotra}) equation with $n_{\sigma}$ and
using the fact that the metric $\tilde{g}_{\mu\nu}$ projects onto the
transverse components, we get

\begin{equation}
  \label{eq:KSIa}
  \mathcal{E}_{e}{}^{m\mu} +2\hat{\mathcal{E}}_{B}{}^{m\mu}
  +120\ell_{\nu}n_{\rho} e^{-2\phi}
  \mathcal{E}_{\tilde{B}}{}^{\alpha_{1}\alpha_{2}\alpha_{3}\mu\nu\rho}
  \Omega_{\alpha_{1}\alpha_{2}\alpha_{3}}{}^{m}
  =
  0\, .
\end{equation}

\noindent
Contracting this identity with $e_{m\, \mu}$,  $n_{\mu}$ and 
$\ell_{\mu}$ and using Eq.~(\ref{eq:KSIotra2}) we get 

\begin{eqnarray}
  \label{eq:KSIb}
  \mathcal{E}_{e}{}^{mn} 
  +120\ell_{\nu}n_{\rho} e^{-2\phi}
  \mathcal{E}_{\tilde{B}}{}^{\alpha_{1}\alpha_{2}\alpha_{3}\mu\nu(m}
  \Omega_{\alpha_{1}\alpha_{2}\alpha_{3}}{}^{n)}
  &  = &
         0\, ,
  \\
  & & \nonumber \\
  \label{eq:KSIc}
   \hat{\mathcal{E}}_{B}{}^{mn}
  & = &
  0\, .
  \\
  & & \nonumber \\
  \label{eq:KSI7}
n_{\mu}\left(  \mathcal{E}_{e}{}^{m\mu}
  +2\hat{\mathcal{E}}_{B}{}^{m\mu}\right)
  & = &
        0\, ,
  \\
  & & \nonumber \\
  \label{eq:KSI4a}
          \ell_{\mu} \left(\mathcal{E}_{e}{}^{m\mu}
  +2\hat{\mathcal{E}}_{B}{}^{m\mu}\right)
  & = &
  0\, .
\end{eqnarray}

\noindent
Combining the last equation with  Eq.~(\ref{eq:KSI4}) we get

\begin{equation}
\label{eq:KSI4c}
  \ell_{\mu} \mathcal{E}_{e}{}^{\mu m}
    =
\ell_{\mu} \hat{\mathcal{E}}_{B}{}^{\mu m}
=
0\, .
\end{equation}

Observe that Eq~(\ref{eq:KSIb}) leads to a relation between the trace of the
spatial components of the Einstein equation and, yet again, the component
$\Omega^{\alpha_{1}\cdots\alpha_{4}}\mathcal{B}_{H\,
  \alpha_{1}\cdots\alpha_{4}}$ of the Bianchi identity. On account of
Eq.~(\ref{eq:KSI3}) we can write

\begin{equation}
\label{eq:KSI6}
\mathcal{E}_{\phi}
=
-\tfrac{1}{2} \mathcal{E}_{e}{}^{m}{}_{m}
=
-60e^{-2\phi}\Omega^{\alpha_{1}\cdots\alpha_{4}}\mathcal{B}_{H\, \alpha_{1}\cdots\alpha_{4}}\, .
\end{equation}

If, instead, we hit Eq.~(\ref{KSI_e_B}) with
$\Gamma_{\mu}\Gamma^{\sigma}\epsilon$ from the right and use
Eq.~(\ref{eq:KSI4}) to eliminate the terms containing $\mathcal{E}_{B}$, we
get
$\ell_{\mu} \mathcal{E}_{e}{}^{\mu  m} =
60\ell_{\mu}\mathcal{E}_{\tilde{B}}{}^{\mu m}$,
from which it follows that

\begin{equation}
  \label{eq:KSI8}
  \ell_{\mu}\mathcal{E}_{\tilde{B}}{}^{\mu m}
  = 
  0\, ,
\end{equation}

Finally, Eq.~(\ref{KSI_e_B}) with
$\Gamma_{\alpha_{1}\cdots\alpha_{4}}\epsilon$ from the right we get

\begin{equation}
  \label{eq:KSI9}
  \begin{aligned}
   \left(e^{a\, \mu}\mathcal{E}_{e\, a}{}^{\nu}
    +2\mathcal{E}_{B}{}^{\mu\nu}\right)
    W_{\mu\alpha_{1}\cdots\alpha_{4}}
    +6
    e^{-2\phi}\mathcal{E}_{\tilde{B}}{}^{\beta_{1}\cdots\beta_{5}\nu}
    \epsilon_{\beta_{1}\cdots\beta_{5}\alpha_{1}\cdots\alpha_{4}\mu}\ell^{\mu}
    & \\
    & \\
    -720 e^{-2\phi}\mathcal{E}_{\tilde{B}\,
    \beta_{1}\beta_{2}\beta_{3}[\alpha_{1}\alpha_{2}}{}^{\nu}
    W^{\beta_{1}\beta_{2}\beta_{3}}{}_{\alpha_{3}\alpha_{4}]}
+720 \ell^{\mu}e^{-2\phi}\mathcal{E}_{\tilde{B}\,  \alpha_{1}\cdots\alpha_{4}\mu}{}^{\nu}
  & =
  0\, .
  \end{aligned}
\end{equation}

\noindent
Decomposing $W$ and contracting the resulting expression with
$n^{\alpha_{4}}$\footnote{Contracting this expression with
  $\Omega^{\alpha_{1}\cdots \alpha_{4}}$ and $\ell^{\alpha_{4}}$ we obtain
  $0=0$ and contracting it with $\ell_{\nu}$ we get an identity that has
  already been derived.}

\begin{equation}
  \label{eq:KSI10}
  \begin{aligned}
    \left(e^{a\, \mu}\mathcal{E}_{e\, a}{}^{\nu}
      +2\mathcal{E}_{B}{}^{\mu\nu}\right)\Omega_{\mu\alpha_{1}\alpha_{2}\alpha_{3}}
    -720\ell_{\mu}n_{\rho}e^{-2\phi}\mathcal{E}_{\tilde{B}\,
      \alpha_{1}\alpha_{2}\alpha_{3}}{}^{\nu\mu\rho}
    & \\
    & \\
    +6e^{-2\phi}\mathcal{E}_{\tilde{B}}{}^{\beta_{1}\cdots\beta_{5}\nu}
    \tilde{\epsilon}_{\beta_{1}\cdots\beta_{5}\alpha_{1}\alpha_{2}\alpha_{3}}
    +1080\ell_{\mu}n_{\rho}e^{-2\phi}
    \mathcal{E}_{\tilde{B}\, \beta_{1}\beta_{2}[\alpha_{1}}{}^{\nu\mu\rho}
    \Omega^{\beta_{1}\beta_{2}}{}_{\alpha_{2}\alpha_{3}]}
    & \\
    & \\
    +720n_{\rho}e^{-2\phi}
    \mathcal{E}_{\tilde{B}\, \beta_{1}\beta_{2}\beta_{3}[\alpha_{1}}{}^{\nu\rho}
    \Omega^{\beta_{1}\beta_{2}\beta_{3}}{}_{\alpha_{2}}\ell_{\alpha_{3}]} 
    -360e^{-2\phi}
    \mathcal{E}_{\tilde{B}\, \beta_{1}\beta_{2}\beta_{3}[\alpha_{1}\alpha_{2}}{}^{\nu}
    \Omega^{\beta_{1}\beta_{2}\beta_{3}}{}_{\alpha_{3}]}
    & =
    0\, .
  \end{aligned}
\end{equation}

This is a complicated identity that we can simplify by hitting it with
$\ell,n,\Omega$ in different ways.

Hitting Eq.~(\ref{eq:KSI10}) with $g^{\alpha_{1}}{}_{\nu}$ and using several
of the identities derived above, we get another constraint on the Bianchi
identity

\begin{equation}
  \label{eq:KSI17}
  20\left(\tilde{g}^{\mu\nu}{}_{\rho\sigma}+\tfrac{1}{4}\Omega^{\mu\nu}{}_{\rho\sigma}\right)
  \mathcal{E}_{\tilde{B}}{}^{\rho\sigma}
    -e^{-2\phi}\mathcal{E}_{\tilde{B}}{}^{\beta_{1}\cdots\beta_{6}}
    \tilde{\epsilon}_{\beta_{1}\cdots\beta_{6}}{}^{\mu\nu}
    =
    0\, .
\end{equation}

To summarize, the components of the equations of motion of the
Vierbein, the Kalb-Ramond 2-form and its dual implied by supersymmetry
are:

\begin{equation}
  \begin{aligned}
    \mathcal{E}_{e}{}^{++} & = 0\, ,\,\,\,\,\,\,\,
    & & & & &
    \\
    \\
    & &
    \mathcal{E}_{e}{}^{+-}+2\hat{\mathcal{E}}_{B}{}^{+-} & = 0\, ,\,\,\,\,
    \\
    \\
    \mathcal{E}_{e}{}^{+m} & = 0\, ,\,\,\,\,\,
    & 
    \mathcal{E}_{B}{}^{+m} & = 0\, ,
    &
    \mathcal{E}_{\tilde{B}}{}^{+m} & = 0\, ,
    &
    \\
    \\
    & &
    \mathcal{E}_{e}{}^{m-}+2\hat{\mathcal{E}}_{\tilde{B}}{}^{m-} & = 0\, ,
    \\
    \\
    \mathcal{E}_{e}{}^{mn} & =
    -120\ell_{\nu}n_{\rho} e^{-2\phi}
  \mathcal{E}_{\tilde{B}}{}^{\alpha_{1}\alpha_{2}\alpha_{3}\mu\nu(m}
  \Omega_{\alpha_{1}\alpha_{2}\alpha_{3}}{}^{n)}\, ,
    &
   \hat{\mathcal{E}}_{B}{}^{mn} & = 0\, ,
    &
    &
    &
      \end{aligned}
\end{equation}

\noindent
and those of the gauge fields and dilaton are

\begin{equation}
  \hat{\mathcal{E}}_{A}{}^{+}  = 0\, ,
  \hspace{1cm}
  \hat{\mathcal{E}}_{A}{}^{m}  = 0\, ,
  \hspace{1cm}
  \mathcal{E}_{\phi} = 60\mathcal{E}_{\tilde{B}}{}^{+-}\, .
\end{equation}

\subsection{Remaining equations for supersymmetric solutions}
\label{sec-EOMS}

A possible choice of independent equations of motion to be checked is

\begin{equation}
  \begin{aligned}
    \mathcal{E}_{e}^{--} & = 0\, ,\,\,\,\,\,
    \\
    \\
    \mathcal{E}_{B}^{+-} & = 0\, ,\,\,\,\,\,
    & 
    \mathcal{E}_{B}^{-m} & = 0\, ,\,\,\,\,\, 
    \\
    \\
    \mathcal{B}_{H\, \mu\nu\rho\sigma} & = 0\, ,\,\,\,\,\,
  \\
  \\
  \hat{\mathcal{E}}_{A}{}^{-} & = 0\, ,
  \end{aligned}
\end{equation}

\noindent
although some combinations of the components of the Bianchi identity are
automatically satisfied for supersymmetric field configurations. It should be
remembered that, in order to derive some of the relations, we had to assume
that the Bianchi identity is satisfied at lowest order in $\alpha'$.

\section{Discussion}
\label{sec-discussion}

In this paper we have re-analyzed the problem of characterizing all the
supersymmetric solutions of the Heterotic Superstring effective action to
first order in $\alpha'$ working in a general spinorial basis, instead of
working on a particular or privileged basis as in it is done in
Refs.~\cite{Gran:2005wf,Gran:2007fu,Gran:2007kh}. 

Thus, we have first computed the algebra of bilinears in an arbitrary basis in
Section~\ref{sec-d10bilinearalgebra}. It is this computation that allowed us
to obtain the conditions necessary for unbroken supersymmetry and to construct
the supersymmetry projectors necessary to prove the sufficiency of the
conditions in an arbitrary basis. As explained below
Eq.~(\ref{eq:Pi-spinorprojector}), the explicit form of the supersymmetry
projector for the Killing spinors suggests a physical interpretation of
minimally supersymmetric field configurations which is entirely lost if one
works in a privileged spinor basis. The form of the projector
Eq.~(\ref{eq:Pi-spinorprojector}), for instance, is general,
basis-independent. One can always evaluate it in a particular basis such as
the one used in Refs.~\cite{Gran:2005wf,Gran:2007fu,Gran:2007kh} if that is
needed in a specific calculation.

Another important result is the derivation of the set of relations existing
between the equations of motion evaluated on supersymmetric configurations to
first order in $\alpha'$ obtained in Section~\ref{sec-KSIs} using the Killing
Spinor Identities (KSIs). One of the novelties in this result is the procedure
through which we have obtained it, using the KSIs in this, more complicated,
context. We have also shown how to include the Bianchi identity of the 3-form
Kalb-Ramond field strengths in the KSIs.

Finally, we have also re-derived the set of conditions necessary for unbroken
supersymmetry, summarized in Section~\ref{sec-d10summary}. They are, of
course, completely equivalent to those obtained in
Refs.~\cite{Gran:2005wf,Gran:2007fu,Gran:2007kh}.

The computation of the bilinear algebra in an arbitrary basis is a very useful
result because the algebras of bilinears of half-maximal
supergravities\footnote{Actually, of any half-maximal supersymmetric theory,
  including global supersymmetry.} in lower dimensions are exactly the same,
up to relabeling of the components of the bilinear forms \cite{kn:FO}, which
implies the existence of a Spin(7) structure hidden in any supersymmetric
solution of any half-maximal theory. This observation deserves further
discussion.\footnote{Observe that, ultimately, this is a property of the
  Clifford algebra itself that will hold whenever (1,9)-dimensional
  Majorana-Weyl spinors are at play. For instance, in Ref.~\cite{kn:CSS} it
  has been shown that any 10-dimensional Lorentzian manifold admits a real
  chiral spinor if and only if it admits a null vector such that the
  associated metric in the corresponding ``screen bundle'' is Spin(7). }

Let us consider the supersymmetry condition of the Yang-Mills fields
Eq.~(\ref{gau_2_reduced-bis}) which we rewrite here for the sake of
convenience:

\begin{equation}
  \label{eq:omegaselfduality}
  F^{A\, mn} = \tfrac{1}{2}\Omega^{mnpq}F^{A}{}_{pq}\, ,
  \hspace{1cm}
  m,n,p,q=1,\cdots, 8\, .
\end{equation}

This equation can be seen as an 8-dimensional generalization of the
4-dimensional self-duality condition

\begin{equation}
  \label{eq:epsilonselfduality}
  F^{A\, mn} = \tfrac{1}{2}\epsilon^{mnpq}F^{A}{}_{pq}\, ,
    \hspace{1cm}
  m,n,p,q=1,\cdots, 4\, ,
\end{equation}

\noindent
that characterizes 4-dimensional Yang-Mills instantons such as the BPST
instanton with gauge group SU(2)~$\subset$~SO(4) \cite{Belavin:1975fg}.  This
instanton is part of the \textit{gauge 5-brane} solution of Heterotic
Supergravity \cite{Strominger:1990et}, sourcing the gravitational and dilaton
field. Combined with the \textit{solitonic (or NS) 5-brane} of
Ref.~\cite{Duff:1990wv} as in Ref.~\cite{Callan:1991dj}, in which the source
is a magnetic Kalb-Ramond field, one can obtain the so-called
\textit{symmetric 5-brane} \cite{Callan:1991dj}, which is considered an exact
solution of the Heterotic Superstring effective action to all orders in
$\alpha'$. It is clear that this solution should fit into our general result
and that the gauge field satisfying Eq.~(\ref{eq:epsilonselfduality}) should
obey Eq.~(\ref{eq:omegaselfduality}) for some Spin(7) structure 4-form. As a
mater of fact, one can view the Spin(7) structure 4-form as a collection of
volume forms in 4-dimensional manifolds (hyper-planes in 8-dimensional Euclidean
space, \cite{kn:Joyce}) and, if we simply restrict
Eq.~(\ref{eq:omegaselfduality}) to the 4-dimensional subspace in which the
gauge field is defined to live, we just get (up to a sign)
Eq.~(\ref{eq:epsilonselfduality}). Therefore,
Eq.~(\ref{eq:epsilonselfduality}) is included as a particular case in
Eq.~(\ref{eq:omegaselfduality}).

A solution to Eq.~(\ref{eq:omegaselfduality}) that does not assume that the
gauge field lives in less than 8 dimensions is the so-called
\textit{octonionic instanton} of Nicolai and Fubini \cite{Fubini:1985jm},
whose gauge group is Spin(7)~$\subset$~SO(8). Observe that the use of a
generic basis for the Spin(7) structure 4-form and the knowledge of the
algebra it satisfies plays an important r\^ole in the construction of the
solution. The octonionic instanton has been used to construct the
\textit{octonionic superstring soliton} of Ref.~\cite{Harvey:1990eg}, which
is, actually, a $\mathcal{O}(\alpha')$ solution of the Heterotic Superstring
effective action preserving exactly one supersymmetry and, therefore, a very
good example of the characterization discussed in this paper. It is, on the
other hand, a solution closely related to that of the symmetric 5-brane
mentioned above: both of them are sourced by Yang-Mills instanton fields
satisfying Eq.~(\ref{eq:omegaselfduality}), the main difference being the
number of transverse directions the gauge fields do not depend upon and the
number of isometries of the metric (8 to 4) and the absence of a (known)
solution of the same kind with no gauge fields, sourced only by the
Kalb-Ramond field, in analogy with the solitonic 5-brane.\footnote{The
  torsionful spin connection $\Omega_{(+)}$ of that solution is also that of
  an ``octonionic instanton''.}

It should be possible to consider solutions to Eq.~(\ref{eq:omegaselfduality})
for cases in between the full 8-dimensional dependence of the octonionic
instanton and the 4-dimensional dependence of the BPST instanton, with gauge
groups that can be embedded in SO$(n)$, $4<n<8$ and which should coincide with
(some of) the special holonomy groups found in
Refs.~\cite{Gran:2005wf,Gran:2007fu,Gran:2007kh}. Let us consider, for
instance, the $n=7$ case, in which the gauge field lives in a 7-dimensional
space or, alternatively, does not depend on one of the original 8 transverse
coordinates, $x^{8}$, say. it is not difficult to see that only the components
$\Omega_{mnpq}$ with $m,n,p,q=1,\dots,7$ occur in
Eq.~(\ref{eq:omegaselfduality}) and, due to the 8-dimensional selfduality of
the Spin(7) structure, they can be rewritten in terms of the 3-form
$\Sigma_{mnp}\equiv \Omega_{mnp8}$ which satisfies the algebra of a G$_{2}$
structure. Thus, it should not be surprising that one can construct G$_{2}$
instantons in $\mathbb{R}^{7}$ using an ansatz similar to Nicolai and Fubini's
(or 't~Hooft's for the BPST instanton) and a complete solution of the
Heterotic Superstring effective action sourced by such an instanton. As a
matter of fact, both the instanton and the solution were constructed in
Ref.~\cite{Gunaydin:1995ku}. It should be possible to find more intermediate
cases, with a number of isometries ranging between 2 (the \textit{string},
Spin(7) case) and 6 (the \textit{fivebrane}, SU(2) case), so they can be
interpreted as $(n+1)$-brane solitons. Their existence would greatly enhance
the spectrum of non-perturbative extended solitons of the Heterotic
Superstring. Work in this direction is in progress.

\section*{Acknowledgments}

We would like to thank C.S.~Shahbazi for many conversations on spinors,
Killing spinors and related subjects and A.~Ruip\'erez for pointing to us an
error in a previous version of this paper and U.~Gran and G.~Papadopoulos for
clarifications on their previous work. This work has been supported in part by
the MCIU, AEI, FEDER (UE) grant PGC2018-095205-B-I00 and by the Spanish
Research Agency (Agencia Estatal de Investigaci\'on) through the grant IFT
Centro de Excelencia Severo Ochoa SEV-2016-0597.  The work of AF has been
partially supported by the Angelo della Riccia Foundation Fellowship and
partially by the German Research Foundation (DFG) via the Emmy Noether program
``Exact Results in Extended Holography''.  TO wishes to thank M.M.~Fern\'andez
for her permanent support.

\appendix
\section{$d=10$ gamma matrices, spinors and the algebra of bilinears}
\label{sec-d10spinorsetc}

\subsection{$d=10$ gamma matrices and spinors}
\label{sec-d10spinors}

In this appendix, $\Gamma^{a}$, $a, b, c, \ldots =0,\cdots,9$ are the
10-dimensional gamma matrices, satisfying the Clifford algebra

\begin{equation}
  \{ \Gamma_{a} , \Gamma_{b} \} = 2 \eta_{ab} \, ,
\end{equation}

\noindent
where $(\eta_{ab})=\mathrm{diag}(+-\cdots-)$ is the 10-dimensional
Minkowski metric.

The chirality matrix $\Gamma_{11}$ is defined to satisfy the relations

\begin{equation}
  \label{eq:d10gammaduality}
  \Gamma_{11}\Gamma^{a_{1}\cdots a_{n}}
  =
  \frac{(-1)^{[(10-n)/2]+1}}{(10-n)!}
  \epsilon^{a_{1}\cdots a_{n}b_{1}\cdots b_{10-n}}
  \Gamma_{b_{1}\cdots b_{10-n}}\, ,
\end{equation}

\noindent
so that, in particular,

\begin{subequations}
  \begin{align}
    \Gamma_{11}
    & =
      \tfrac{1}{10!}\epsilon_{b_{1}\cdots b_{10}} \Gamma^{b_{1}\cdots b_{10}}
      =
      -\Gamma^{0}\cdots \Gamma^{9}\, ,
    \\
    \nonumber \\
  \Gamma^{a_{1}\cdots a_{5}}\Gamma_{11}
  & =
  \tfrac{1}{5!}
  \epsilon^{a_{1}\cdots a_{5}b_{1}\cdots b_{5}}
  \Gamma_{b_{1}\cdots b_{5}}\, ,
  \end{align}
\end{subequations}

\noindent
where $\epsilon^{0\cdots 9}= - \epsilon_{0\cdots 9}=+1$.
Furthermore,

\begin{equation}
  \Gamma_{11}^{\dagger}
  =
  \Gamma_{11}^{T}
  =
  \Gamma_{11}\, .
\end{equation}

The charge-conjugation and Dirac-conjugation matrices
$\mathcal{C}$ and $\mathcal{D}$ are defined by the properties

\begin{subequations}
  \begin{align}
    \label{eq:Dirac10}
    \mathcal{D}\Gamma^{a}
    & =
      \Gamma^{a\, \dagger}\mathcal{D}\, ,
      \,\,\,\,\,\,
      \Rightarrow
      \,\,\,\,\,\,
      \mathcal{D}\Gamma^{a_{1}\cdots a_{n}}
     =
      (-1)^{[n/2]}\left(\Gamma^{a_{1}\cdots a_{n}}\right)^{\dagger}\mathcal{D}\, ,
    \\
    \nonumber \\
    \label{eq:charge10}
    \mathcal{C}\Gamma^{a}
    & =
      \Gamma^{a\, T}\mathcal{C}\, ,
      \,\,\,\,\,\,
      \Rightarrow
      \,\,\,\,\,\,
      \mathcal{C}\Gamma^{a_{1}\cdots a_{n}}
      =
      (-1)^{[n/2]}\left(\Gamma^{a_{1}\cdots a_{n}}\right)^{T}\mathcal{C}\, .
  \end{align}
\end{subequations}

\noindent
The particular matrices we have chosen are

\begin{equation}
  \mathcal{C} = -i \Gamma^{0} \Gamma^{3} \Gamma^{4} \Gamma^{6} \Gamma^{8}\, ,
  \hspace{1cm}
  \mathcal{D} =  \Gamma^{0} \ , 
\end{equation}

\noindent
and satisfy

\begin{subequations}
  \begin{align}
    \mathcal{C}^{T}
    & =
      \mathcal{C}
      =
      -\mathcal{C}^{-1}
      =
      -\mathcal{C}^{\dagger}\, ,
    \\
    \nonumber \\
    \mathcal{D}^{\dagger}
    & =
      \mathcal{D}\, ,
    \\
    \nonumber \\
    \mathcal{D}\Gamma_{11}
    & =
    -\Gamma_{11}\mathcal{D}\, ,
    \\
    \nonumber \\
    \mathcal{C}\Gamma_{11}
    & =
    -\Gamma_{11}\mathcal{C}\, .
  \end{align}
\end{subequations}

Given a 10-dimensional spinor $\psi$, using these matrices, we define
its Dirac and Majorana conjugates, respectively $\bar{\psi}$ and
$\psi^{c}$, by

\begin{subequations}
  \begin{align}
    \bar{\psi}
    & \equiv
      \psi^{\dagger}\mathcal{D}\, ,
    \\
    \nonumber \\
    \psi^{c}
    & \equiv
      \psi^{T}\mathcal{C}\, .
  \end{align}
\end{subequations}

\noindent
Majorana spinors are defined by the property

\begin{equation}
\bar{\psi} = \psi^{c}\, .  
\end{equation}

\noindent
With the particular choices of $\mathcal{C}$ and $\mathcal{D}$ that we have
made, they are neither purely real nor purely imaginary, but this is the most
convenient choice for reducing them to symplectic-Majorana spinors in $d=6$
dimensions (which will be useful in a forthcoming work \cite{kn:FO}).

The supersymmetry parameter of Heterotic Supergravity, $\epsilon$,
is a Majorana-Weyl spinor. We choose the convention

\begin{equation}
  \Gamma_{11}\epsilon = +\epsilon\, ,
  \,\,\,\,\,
  \Rightarrow
  \,\,\,\,\,
  \bar{\epsilon}\Gamma_{11} = -\bar{\epsilon}\, .
\end{equation}

\subsection{$d=10$ spinor bilinears}
\label{sec-d10bilinears}

Let us consider the bilinears of these spinors (taken as commuting)
$\bar{\epsilon}\Gamma^{a_{1}\cdots a_{n}}\epsilon$. Using the above
properties we find

\begin{subequations}
  \begin{align}
    \bar{\epsilon}\Gamma^{a_{1}\cdots a_{n}}\epsilon
    & =
      0\, ,
      \,\,\,\,\,\, \forall n\,\,\, \text{even}\, .
    \\
    \nonumber \\
    \left(\bar{\epsilon}\Gamma^{a_{1}\cdots a_{n}}\epsilon\right)^{T}
    & =
      (-1)^{[n/2]}\, \bar{\epsilon}\Gamma^{a_{1}\cdots a_{n}}\epsilon\, ,
    \\
    \nonumber \\
    \left(\bar{\epsilon}\Gamma^{a_{1}\cdots a_{n}}\epsilon\right)^{\dagger}
    & =
      (-1)^{[n/2]}\, \bar{\epsilon}\Gamma^{a_{1}\cdots a_{n}}\epsilon\, ,
  \end{align}
\end{subequations}

\noindent
from which it follows that only for $n=1 (\text{mod}\, 4 )$ the
bilinear is generically non-vanishing and that, in those cases, it is
real. Since the $n$ and $\tilde{n}=10-n$ bilinears are related by the duality
Eq.~(\ref{eq:d10gammaduality}) we end up with just two independent,
real bilinears: a 1-form that we denote by $\ell_{a}$ and a selfdual
5-form that we denote by $W_{a_{1}\cdots a_{5}}$

\begin{subequations}
  \begin{align}
    \ell_{a}
    & \equiv
      \bar{\epsilon}\Gamma_{a}\epsilon\, ,
    \\
    \nonumber \\
    W_{a_{1}\cdots a_{5}}
    & \equiv 
      \bar{\epsilon}\Gamma_{a_{1}\cdots a_{5}}\epsilon
      =
      \tfrac{1}{5!}
      \epsilon^{a_{1}\cdots a_{5}b_{1}\cdots b_{5}}
      W_{b_{1}\cdots b_{5}}\, .
  \end{align}
\end{subequations}

\subsection{$d=10$ Fierz identities}
\label{sec-d10Fierzidentities}

If $\{ \mathcal{O}^{I} \}$ and $\{ \mathcal{O}_{I} \}$ are dual bases
of $32\times 32$ matrices with the trace as inner product 

\begin{equation}
\mathrm{Tr}(\mathcal{O}^{I}\mathcal{O}_{J})= 32\delta^{I}{}_{J}\, ,  
\end{equation}

\noindent
$M$ and $N$ are two arbitrary $32\times 32$ matrices and
$\lambda,\chi,\psi,\varphi$ are four 32-component commuting spinors,
the Fierz identities take the form

\begin{equation}
  (\bar{\lambda}M\chi)(\bar{\psi}N\varphi)
  =
  \frac{1}{32} \sum_{I}(\bar{\lambda}M\mathcal{O}^{I}N\varphi)
  (\bar{\psi}\mathcal{O}_{I}\chi)\, .
\end{equation}

Choosing the dual bases in the space of $32\times 32$ matrices

\begin{subequations}
  \begin{align}
    \{ \mathcal{O}^{I} \}
    & \equiv
      \{ \mathds{1}, \Gamma^{a}, i\Gamma^{ab}, i\Gamma^{abc},
      \Gamma^{a_{1}\cdots a_{4}}, \Gamma^{a_{1}\cdots a_{5}},
      \Gamma^{a_{1}\cdots a_{4}}\Gamma_{11},  \Gamma^{abc}\Gamma_{11},
      i\Gamma^{ab}\Gamma_{11}, i\Gamma^{a}\Gamma_{11}, \Gamma_{11}
      \}\, ,
    \\
    \nonumber \\
    \{ \mathcal{O}_{I} \}
    & \equiv
      \{ \mathds{1}, \Gamma_{a}, i\Gamma_{ab}, i\Gamma_{abc},
      \Gamma_{a_{1}\cdots a_{4}}, \Gamma_{a_{1}\cdots a_{5}},
      \Gamma_{a_{1}\cdots a_{4}}\Gamma_{11},  \Gamma_{abc}\Gamma_{11},
      i\Gamma_{ab}\Gamma_{11}, i\Gamma_{a}\Gamma_{11}, \Gamma_{11}
      \}\, ,
  \end{align}
\end{subequations}

\noindent
and assuming that the spinors $\lambda,\chi,\psi,\varphi$ have
positive chirality (so $M$ and $N$ have to be products of odd numbers
of gammas and, therefore, anticommute with $\Gamma_{11}$), the Fierz
identity takes the explicit form

\begin{align}
  (\bar{\lambda}M\chi)(\bar{\psi}N\varphi)
  & =
    \frac{1}{16}
    (\bar{\lambda}M\Gamma^{a}N\varphi)
  (\bar{\psi}\Gamma_{a}\chi)
    -\frac{1}{16\cdot 3!}
    (\bar{\lambda}M\Gamma^{abc}N\varphi)
    (\bar{\psi}\Gamma_{abc}\chi)
    \nonumber \\
  \nonumber \\
  &
    +\frac{1}{32\cdot 5!}
    (\bar{\lambda}M\Gamma^{a_{1}\cdots a_{5}}N\varphi)
  (\bar{\psi}\Gamma_{a_{1}\cdots a_{5}}\chi)\, .
\end{align}

Finally, if $\lambda=\chi=\psi=\varphi=\epsilon$, the 3-form vanishes
and, using the above definitions for the bilinears, we are left with
just

\begin{equation}
  (\bar{\epsilon}M\epsilon)(\bar{\epsilon}N\epsilon)
   =
    \frac{1}{16}
    (\bar{\epsilon}M\Gamma^{a}N\epsilon)\ell_{a}
    +\frac{1}{32\cdot 5!}
    (\bar{\epsilon}M\Gamma^{a_{1}\cdots a_{5}}N\epsilon)
    W_{a_{1}\cdots a_{5}}\, ,
\end{equation}

\subsection{$d=10$ bilinear algebra}
\label{sec-d10bilinearalgebra}

It is now straightforward to compute the products of the bilinears
$\ell_{a}$ and $W_{a_{1}\cdots a_{5}}$ using the Fierz identities we
just derived in the previous section. Each of the identities obtained
has been checked to be consistent with the self-duality of the 5-form,
which imposes strong constraints on the possible combinations that can
occur in the right-hand side.

To start with, we find for the product of 1-forms 

\begin{equation}
\label{1_{1}_bilinear}
\ell_{a} \ell_{b}
=
\frac{1}{14 \cdot 4!} W_{a}{}^{c_{1} ... c_{4}} W_{b c_{1} ... c_{4}} \, , 
\end{equation}

\noindent
and we observe that the right-hand-side is equal to itself when we
replace $W$ by its dual, in agreement with the invariance under
duality of the left-hand side.

For the product of a 1-form and one self-dual 5-form we have
obtained\footnote{Here and in what follows, it is assumed that all 
  indices that have the same letter ($b_{1},\cdots, b_{5}$ in this
  case) are always antisymmetrized. The brackets of the
  antisymmetrizers have been suppressed to avoid the cluttering of the
  formulae.}

\begin{equation}
\label{1_5_bilinear}
\ell^{a} W_{b_{1} \cdots b_{5}}
=
\tfrac{5}{7} \ell_{b_{1}} W_{b_{2} \cdots b_{5}}{}^{a}
-\tfrac{5}{14} W_{b_{1}b_{2}b_{3}}{}^{c_{1} c_{2}} W_{b_{4} b_{5} c_{1} c_{2}}{}^{a}\, , 
\end{equation}

\noindent
where we have used the fact that the term 

\begin{equation}
\eta^{a}{}_{b_{1}} W_{b_{2} \cdots b_{5}d} \ell^{d}\, ,  
\end{equation}

\noindent
which occurs in the right-hand side vanishes identically due to the
self-duality of $W$. The expression in the right-hand side is self-dual
in the five $b_{i}$ indices, just as the left-hand side.

In order to express the product of two self-dual 5-forms we have
defined, first, for the sake of convenience, the following products and
contractions of the self-dual 5-form with itself:

\begin{subequations}
\begin{align}
  A
  & \equiv
    W_{a_{1}\cdots a_{4}b_{5}} W_{b_{1}\cdots b_{4}a_{5}}\, ,
  \\
  \nonumber \\
  B
  & \equiv
    W_{a_{1}a_{2} a_{3}b_{4}b_{5}} W_{b_{1}b_{2}b_{3} a_{4}a_{5}}\, ,
  \\
  \nonumber \\
  C
  & \equiv
    W_{a_{1}a_{2} a_{3}}{}^{c_{1}c_{2}} W_{b_{1}b_{2}b_{3} c_{1}c_{2}}
    \eta_{a_{4}a_{5},\, b_{4}b_{5}}\, ,
  \\
  \nonumber \\
  D
  & \equiv
    W_{a_{1}a_{2} b_{3}}{}^{c_{1}c_{2}} W_{b_{1}b_{2}a_{3} c_{1}c_{2}}
    \eta_{a_{4}a_{5},\, b_{4}b_{5}}\, ,
  \\
  \nonumber \\
  E
  & \equiv
    W_{a_{1}}{}^{c_{1}\cdots c_{4}} W_{b_{1}c_{1}\cdots c_{4}}
    \eta_{a_{2}\cdots a_{5},\, b_{2}\cdots b_{5}}\, ,
  \\
  \nonumber \\
  F
  & \equiv
    W_{a_{1}\cdots a_{5}} W_{b_{1}\cdots b_{5}}\, .
\end{align}
\end{subequations}

\noindent
The rest of the terms quadratic in $W$ that can occur in the
right-hand side are linear combinations of them, as can be seen by
replacing $W$ by its dual. These relations are

\begin{subequations}
  \begin{align}
    W_{a_{1}a_{2}}{}^{c_{1}c_{2}c_{3}} W_{b_{1}b_{2} c_{1}c_{2} c_{3}}
    \eta_{a_{3}a_{4}a_{5},\, b_{3}b_{4}b_{5}}
    & =
      \tfrac{1}{2}E\, ,
    \\
    \nonumber \\
    W_{a_{1}b_{2}}{}^{c_{1}c_{2}c_{3}} W_{b_{1}a_{2} c_{1}c_{2} c_{3}}
    \eta_{a_{3}a_{4}a_{5},\, b_{3}b_{4}b_{5}}
    & =
      -\tfrac{1}{4}E\, ,
    \\
    \nonumber \\
    W_{a_{1}\cdots a_{4}}{}^{c} W_{b_{1}\cdots b_{4}c}
    \eta_{a_{5},\, b_{5}}
    & =
      4C-E\, ,
    \\
    \nonumber \\
    W_{a_{1}a_{2}a_{3} b_{4}}{}^{c} W_{b_{1}b_{2}b_{3}a_{4}c}
    \eta_{a_{5},\, b_{5}}
    & =
      \tfrac{1}{4}C+\tfrac{9}{4}D-\tfrac{1}{4}E\, ,
    \\
    \nonumber \\
    W_{a_{1}a_{2}b_{3} b_{4}}{}^{c} W_{b_{1}b_{2}a_{3}a_{4}c}
    \eta_{a_{5},\, b_{5}}
    & =
      2D-\tfrac{1}{6}E\, ,
  \end{align}
\end{subequations}

\noindent
and they allow us to use $A,B,C,D$ and $E$ as a basis for these
products.

Then, using this notation, we find

\begin{align}
\label{5_5_bilinear}
  W_{a_{1} \cdots a_{5}} W_{b_{1} \cdots b_{5}}
  & =
    \tfrac{1}{3}\, \left(
    \ell_{a_{1}} \epsilon_{a_{2} \cdots a_{5} b_{1} \cdots b_{5} d} \ell^{d}
    + a\leftrightarrow b \right)
    \nonumber \\
  \nonumber \\
  &
    -80\,
    \left(\ell_{a_{1}} W_{a_{2}a_{3}b_{1}b_{2}b_{3}} \eta_{a_{4}a_{5},\, b_{4}b_{5}}
    + a\leftrightarrow b \right)
    \nonumber \\
  \nonumber \\
  &
    +80 \, \ell_{a_{1}} \ell_{b_{1}} \eta_{a_{2} \cdots a_{5},\, b_{2} \cdots b_{5}}
  \nonumber \\
  \nonumber \\
  &
    +\tfrac{5}{3}A +\tfrac{20}{3}B -\tfrac{20}{3}C -60 D +\tfrac{25}{3}E\, .
\end{align}

\noindent
All  terms in the right-hand side of this expression are
$a\leftrightarrow b$ symmetric, as the left-hand side.

Furthermore, the left-hand side is self-dual in the $a$ and $b$
indices separately. It can be checked that the right-hand side has the
same property: the combination of the first three terms is self-dual
and the combination of terms in the fourth line is also
self-dual,\footnote{Actually, up to a global factor, it is the only
  self-dual combination.}  as can be seen by using the properties

\begin{subequations}
  \begin{align}
    \star A
    & =
      \tfrac{1}{5}F -4C +E\, ,
    \\
    \nonumber \\
    \star B
    & =
      \tfrac{1}{10}F -3C+E\, ,
          \\
    \nonumber \\
    \star C
    & =
      \tfrac{1}{10}F -A+B\, ,
          \\
    \nonumber \\
    \star D
    & =
      \tfrac{1}{30}F -\tfrac{1}{3}A +\tfrac{1}{3}B -\tfrac{1}{3}C +D\, ,
    \\
    \nonumber \\
    \star E
    & =
      \tfrac{1}{5}F -3A +4B\, .
  \end{align}
\end{subequations}

\subsubsection{Consequences}

The selfduality of $W$ implies

\begin{equation}
W^{2} \equiv W^{a_{1}\cdots a_{5}} W_{a_{1}\cdots a_{5}}=0  \, .
\end{equation}

\noindent
Then, Eq.~(\ref{1_{1}_bilinear}) implies that $\ell$ is null:

\begin{equation}
\ell^{2}=\ell^{a}\ell_{a}=0\, .  
\end{equation}

\noindent
Lowering the upper index of 
Eq.~(\ref{1_5_bilinear})  and antisymmetrizing it with the rest leads to

\begin{equation}
\ell_{b_{1}}W_{b_{2}\cdots b_{6}} =0\, ,
\end{equation}

\noindent
which implies that

\begin{equation}
\label{eq:WlO}
  W_{a_{1}\cdots a_{5}} = 5 \ell_{a_{1}}\Omega_{a_{2}\cdots a_{5}}\, ,
\end{equation}

\noindent
for a certain 4-form $\Omega$. We will see that this 4-form, which was first
found in the supergravity context in Ref.~\cite{deWit:1983gs}, satisfies the
relations of a Spin(7) structure. Plugging this result back into
Eq.~(\ref{1_5_bilinear}) and contracting now the upper index with one of the
lower ones we find that

\begin{equation}
  \ell^{b}W_{a_{1}\cdots a_{4}b}=0\, ,
  \,\,\,\,\,
  \Rightarrow
  \,\,\,\,\,
  \ell^{b}\Omega_{a_{1}a_{2}a_{3} b}=0\, ,
\end{equation}

\noindent
so $\Omega$ lives in the 8-dimensional space transverse to the null vector $\ell$.
It is useful to introduce a null vector $n$ dual to $\ell$:

\begin{equation}
  n^{2}=0\, ,
  \hspace{1cm}
  n^{a}\ell_{a} =1\, ,
\end{equation}

\noindent
and define the metric induced in the 8-dimensional space transverse to $\ell$ as

\begin{equation}
  \label{eq:inducedmetric}
  \tilde{\eta}_{ab}
  \equiv
  \eta_{ab}-2\ell_{(a}n_{b)}\, ,
\end{equation}

\noindent
and a fully antisymmetric tensor in that space as well

\begin{equation}
  \tilde{\epsilon}^{c_{1}\cdots c_{8}}
  \equiv
  \epsilon^{abc_{1}\cdots c_{8}}n_{a}\ell_{b}\, ,
  \hspace{1cm}
  \tilde{\epsilon}_{c_{1}\cdots c_{8}}
  \equiv
  \epsilon_{abc_{1}\cdots c_{8}}n^{a}\ell^{b}\, .
\end{equation}

\noindent
$\Omega$ satisfies

\begin{subequations}
  \begin{align}
    \label{eq:Oselfdual}
    \Omega_{a_{1}\cdots a_{4}}
    & =
      \tilde{\eta}_{a_{1}\cdots a_{4}}{}^{b_{1}\cdots b_{4}} \Omega_{b_{1}\cdots b_{4}}\, ,
    \\
    \nonumber \\
      \Omega^{a_{1}\cdots a_{4}}
    & =
      \tfrac{1}{4!} \tilde{\epsilon}^{a_{1}\cdots a_{4}b_{1}\cdots b_{4}}
      \Omega_{b_{1}\cdots b_{4}}\, .
  \end{align}
\end{subequations}

\noindent
Furthermore, using Eq.~(\ref{eq:WlO}) in Eqs.~(\ref{1_{1}_bilinear}) and
Eq.~(\ref{1_5_bilinear}), we find

\begin{subequations}
  \begin{align}
    \label{eq:O2}
    \Omega_{a_{1}\cdots a_{4}}  \Omega^{a_{1}\cdots a_{4}}
    & \equiv
      \Omega^{2}
       = 14\cdot 4!\, ,
    \\
    \nonumber \\
    \label{eq:O2O2}
    \Omega_{a_{1}a_{2}}{}^{b_{1}b_{2}} \Omega_{a_{3}a_{4}b_{1}b_{2}}
    & =
      -4 \Omega_{a_{1}\cdots a_{4}}\, .
  \end{align}
\end{subequations}

The selfduality of $\Omega$ in the 8-dimensional transverse space
Eq.~(\ref{eq:Oselfdual}) together with Eq.~(\ref{eq:O2}) implies

\begin{subequations}
  \begin{align}
    \label{eq:O1O1}
    \Omega_{a_{1}a_{2}a_{3}}{}^{c}\Omega_{b_{1}b_{2}b_{3}c}
    & =
      -21\tilde{\eta}_{a_{1}a_{2}a_{3},\, b_{1}b_{2}b_{3}}
      +\tfrac{9}{4}\Omega_{a_{1}a_{2}}{}^{c_{1}c_{2}}
      \Omega_{b_{1}b_{2}c_{1}c_{2}}\tilde{\eta}_{a_{3}b_{3}}\, ,
    \\
    \nonumber \\
    \label{eq:O3O3eta}
       \Omega_{ac_{1}c_{2}c_{3}}\Omega_{b}{}^{c_{1}c_{2}c_{3}}
    & =
    42\tilde{\eta}_{ab}\, .
  \end{align}
\end{subequations}

The last product, Eq.~(\ref{5_5_bilinear}), gives the following
expression for the product of two 4-forms:

\begin{equation}
 \label{eq:O4O4}
  \begin{array}{rcl}
    \Omega_{a_{1}\cdots a_{4}} \Omega_{b_{1}\cdots b_{4}}
    & = &
      \tfrac{1}{7}  \tilde{\epsilon}_{a_{1}\cdots a_{4}b_{1}\cdots b_{4}}
      +\tfrac{864}{7}  \tilde{\eta}_{a_{1}\cdots a_{4},\, b_{1}\cdots b_{4}}
      -\tfrac{144}{7} \Omega_{a_{1}a_{2}b_{1}b_{2}}\tilde{\eta}_{a_{3}a_{4},\, b_{3}b_{4}}
       \\
    &  \\
    & &
      +\tfrac{16}{7} \Omega_{a_{1}a_{2}a_{3}b_{4}} \Omega_{b_{1}b_{2}b_{3}a_{4}}
      +\tfrac{18}{7} \Omega_{a_{1}a_{2}b_{3}b_{4}} \Omega_{b_{1}b_{2}a_{3}a_{4}}
    \\
    & & \\
    & &
      -\tfrac{36}{7}\Omega_{a_{1}a_{2}}{}^{c_{1}c_{2}} \Omega_{b_{1}b_{2}c_{1}c_{2}}
      \tilde{\eta}_{a_{3}a_{4},\, b_{3}b_{4}}
      +\tfrac{72}{7}\Omega_{a_{1}b_{2}}{}^{c_{1}c_{2}} \Omega_{b_{1}a_{2}c_{1}c_{2}}
      \tilde{\eta}_{a_{3}a_{4},\, b_{3}b_{4}}\, .
  \end{array}
\end{equation}

This result is $a\leftrightarrow b$ symmetric and fully consistent with
Eqs.~(\ref{eq:O2}) and (\ref{eq:O2O2}). 
It immediately leads to these two identities,

\begin{subequations}
  \begin{align}
        \Omega_{a_{1}\cdots a_{4}} \Omega_{a_{5}\cdots a_{8}}
    & =
      \tfrac{1}{5} \tilde{\epsilon}_{a_{1}\cdots a_{8}}\, ,
    \\
    \nonumber \\
    \label{eq:O2O2noantig}
    \Omega_{a_{1}a_{2}}{}^{c_{1}c_{2}} \Omega_{b_{1}b_{2}c_{1}c_{2}}
    & =
   -\tfrac{2}{5} \Omega_{a_{1}b_{2}}{}^{c_{1}c_{2}} \Omega_{b_{1}a_{2}c_{1}c_{2}}
      -\tfrac{12}{5} \Omega_{a_{1}a_{2}b_{1}b_{2}}
      -\tfrac{72}{5} \tilde{\eta}_{a_{1}a_{2},\, b_{1}b_{2}}\, .
  \end{align}
\end{subequations}

Eqs.~(\ref{eq:O2}) and (\ref{eq:O2O2}) can be obtained from the last
of these equations.

Antisymmetrizing Eq.~(\ref{eq:O2O2noantig}), we find 

\begin{subequations}
\begin{align}
\label{eq:O2O2noantisimple}
    \Omega_{a_{1}a_{2}}{}^{c_{1}c_{2}} \Omega_{b_{1}b_{2}c_{1}c_{2}}
    & =
      -4\Omega_{a_{1}a_{2}b_{1}b_{2}}
      +12\tilde{\eta}_{a_{1}a_{2},\, b_{1} b_{2}}\, ,
  \\
  \nonumber \\
      \label{eq:O2O2antisimple}
    \Omega_{a_{1}b_{2}}{}^{c_{1}c_{2}} \Omega_{b_{1}a_{2}c_{1}c_{2}}
   & =
      +4\Omega_{a_{1}a_{2}b_{1}b_{2}}
      +6\tilde{\eta}_{a_{1}a_{2},\, b_{1} b_{2}}\, .
  \end{align}
\end{subequations}

\noindent
Substituting these two identities back  into Eq.~(\ref{eq:O4O4}) we get

\begin{equation}
 \label{eq:O4O4simple}
  \begin{array}{rcl}
    \Omega_{a_{1}\cdots a_{4}} \Omega_{b_{1}\cdots b_{4}}
    & = &
      \tfrac{1}{7}  \tilde{\epsilon}_{a_{1}\cdots a_{4}b_{1}\cdots b_{4}}
      -\tfrac{188}{7} \Omega_{a_{1}a_{2}b_{1}b_{2}}\tilde{\eta}_{a_{3}a_{4},\, b_{3}b_{4}}
       \\
    & & \\
    & &
      +\tfrac{16}{7} \Omega_{a_{1}a_{2}a_{3}b_{4}} \Omega_{b_{1}b_{2}b_{3}a_{4}}
      +\tfrac{18}{7} \Omega_{a_{1}a_{2}b_{3}b_{4}}
        \Omega_{b_{1}b_{2}a_{3}a_{4}}\, .
  \end{array}
\end{equation}

Further simplifications of this general formula are possible, but we will not
try to obtain them here.

Summarizing, the main relations involving the product and contractions
of two 4-forms that we will use are\footnote{Not all these equations
  are independent. We quote all of them for their usefulness.}

\begin{subequations}
  \begin{align}
    \label{eq:Oselfduality}
    \Omega_{a_{1}\cdots a_{4}}
    & =
      \tfrac{1}{4!}\tilde{\epsilon}_{a_{1}\cdots a_{4}b_{1}\cdots b_{4}}
      \Omega^{b_{1}\cdots b_{4}}\, ,
    \\
    &   \nonumber \\
    \label{eq:O4wedgeO4}
    \Omega_{a_{1}\cdots a_{4}} \Omega_{a_{5}\cdots a_{8}}
    & =
      \tfrac{1}{5} \tilde{\epsilon}_{a_{1}\cdots a_{8}}\, ,
    \\
    &   \nonumber \\
   \label{eq:O1O1simple}
    \Omega_{a_{1}a_{2}a_{3}}{}^{c}\Omega_{b_{1}b_{2}b_{3}c}
    & =
      -9\Omega_{a_{1}a_{2}b_{1}b_{2}}\tilde{\eta}_{a_{3}b_{3}}
      +6\tilde{\eta}_{a_{1}a_{2}a_{3},\, b_{1}b_{2}b_{3}}\, ,
    \\
    \nonumber \\
    \label{eq:O2O2noantisimplebis}
    \Omega_{a_{1}a_{2}}{}^{c_{1}c_{2}} \Omega_{b_{1}b_{2}c_{1}c_{2}}
    & =
      -4\Omega_{a_{1}a_{2}b_{1}b_{2}}
      +12\tilde{\eta}_{a_{1}a_{2},\, b_{1} b_{2}}\, ,
  \\
  \nonumber \\
      \label{eq:O2O2antisimplebis}
    \Omega_{a_{1}b_{2}}{}^{c_{1}c_{2}} \Omega_{b_{1}a_{2}c_{1}c_{2}}
   & =
      +4\Omega_{a_{1}a_{2}b_{1}b_{2}}
      +6\tilde{\eta}_{a_{1}a_{2},\, b_{1} b_{2}}\, .
    \\
    \nonumber \\
    \label{eq:O2O2bis}
    \Omega_{a_{1}a_{2}}{}^{b_{1}b_{2}} \Omega_{a_{3}a_{4}b_{1}b_{2}}
    & =
      -4 \Omega_{a_{1}\cdots a_{4}}\, ,
    \\
    \nonumber \\
    \label{eq:O3O3}
    \Omega_{ac_{1}c_{2}c_{3}}\Omega_{b}{}^{c_{1}c_{2}c_{3}}
    & =
    42\tilde{\eta}_{ab}\, ,
    \\
    \nonumber \\
      \label{eq:O2bis}
    \Omega_{a_{1}\cdots a_{4}}  \Omega^{a_{1}\cdots a_{4}}
    & \equiv
      \Omega^{2}
       = 14\cdot 4!\, .
\end{align}
\end{subequations}

\subsection{Projectors}
\label{sec-projectors}

The 4-form $\Omega_{m_{1}\cdots m_{4}}$ can be used to construct
projectors acting on 2- and 3-forms in the 8-dimensional transverse
space and on spinors. 

For 2-forms, and 3-forms, respectively we have these two complementary
pairs of projectors\footnote{They are mutually orthogonal and their
  sum is the identity in the space of 2- and 3-forms in 8 dimensions. They
  are properly normalized to be idempotent.}

\begin{subequations}
  \begin{align}
    \Pi^{(+)\, m_{1}m_{2}}{}_{n_{1}n_{2}}
    & =
      \tfrac{3}{4}\left(\tilde{\eta}^{m_{1}m_{2}}{}_{n_{1}n_{2}}
      +\tfrac{1}{6}\Omega^{m_{1}m_{2}}{}_{n_{1}n_{2}}\right)\, ,
    \\
    \nonumber \\
    \Pi^{(-)\, m_{1}m_{2}}{}_{n_{1}n_{2}}
    & \equiv
      \tfrac{1}{4}\left(\tilde{\eta}^{m_{1}m_{2}}{}_{n_{1}n_{2}}
      -\tfrac{1}{2}\Omega^{m_{1}m_{2}}{}_{n_{1}n_{2}}\right)\, ,
    \\
    \nonumber \\
    \Pi^{(+)\,  m_{1}m_{2}m_{3}}{}_{n_{1}n_{2}n_{3}}
    & \equiv
      \tfrac{6}{7}\left(\tilde{\eta}^{m_{1}m_{2}m_{3}}{}_{n_{1}n_{2}n_{3}}
      +\tfrac{1}{4}\Omega^{m_{1}m_{2}}{}_{n_{1}n_{2}}\tilde{\eta}^{m_{3}}{}_{n_{3}}
      \right)\, ,
    \\
    \nonumber \\
    \Pi^{(-)\, m_{1}m_{2}m_{3}}{}_{n_{1}n_{2}n_{3}}
    & \equiv
      \tfrac{1}{7}\left(\tilde{\eta}^{m_{1}m_{2}m_{3}}{}_{n_{1}n_{2}n_{3}}
      -\tfrac{3}{2}\Omega^{m_{1}m_{2}}{}_{n_{1}n_{2}}\tilde{\eta}^{m_{3}}{}_{n_{3}}
      \right)\, .
  \end{align}
\end{subequations}

With them we can decompose 2-forms and 3-forms as follows:

\begin{subequations}
  \begin{align}
    F_{mn}
    & =
      F^{(+)}_{mn}+F^{(-)}{}_{mn}\, ,
    \\
    \nonumber \\
    H_{mnp}
    & =
      H^{(+)}_{mnp}+H^{(-)}{}_{mnp}\, ,
  \end{align}
\end{subequations}

\noindent
where the components satisfy

\begin{equation}
  \begin{array}{rclrcl}
    \Pi^{(\pm)}F^{(\pm)} & = & F^{(\pm)}\, ,& \Pi^{(\pm)}F^{(\mp)} & = & 0\, , \\
    & & & & & \\
    \Pi^{(\pm)}H^{(\pm)} & = & H^{(\pm)}\, ,& \Pi^{(\pm)}H^{(\mp)} & = & 0\, . \\
  \end{array}
\end{equation}

On positive chirality spinors satisfying the constraint

\begin{equation}
\ell_{a}\Gamma^{a}\epsilon\equiv \Gamma^{+} \epsilon =0\, ,
\end{equation}

\noindent
it is consistent to define the projectors

\begin{subequations}
  \begin{align}
    \Pi^{(+)}
    & \equiv
      \tfrac{1}{8}\left(1
      +\tfrac{1}{48}\Omega_{m_{1}\cdots m_{4}}\Gamma^{m_{1}\cdots m_{4}}\right)\, ,
    \\
    \nonumber \\
    \label{eq:Pi-spinorprojector}
       \Pi^{(-)}
    & \equiv
      \tfrac{7}{8}\left(1
      -\tfrac{1}{336}\Omega_{m_{1}\cdots m_{4}}\Gamma^{m_{1}\cdots m_{4}}\right)\, .
  \end{align}
\end{subequations}

Observe that these projectors can be interpreted as supersymmetry projectors
associated to a multiple intersection of S5-branes: if $\omega_{6789}$ is the
volume 4-form of the space transverse to a S5-brane whose worldvolume occupies
the directions $0,1,\cdots,5$, its associate supersymmetry projector is
$\sim (1+\omega_{6789}\Gamma^{6789})$. There is a basis \cite{kn:Joyce} in
which the Spin(7) structure 4-form is a linear combination of 4-forms
associated to the transverse space of intersecting S5-branes and, therefore,
the above projector can be understood as a superposition of the associated
supersymmetric projectors of those S5-branes.  This suggests that the
supersymmetric configurations with minimal supersymmetry can be understood as
configurations describing multiple intersection S5-branes.

The projectors acting on spinors and forms satisfy several
relations. The two relations that we will use are

\begin{subequations}
  \begin{align}
    \label{eq:F+property}
    F^{(+)}_{mn}\Gamma^{mn}\epsilon
    & =
      F^{(+)}_{mn}\Gamma^{mn}\Pi^{(-)}\epsilon\, ,
    \\
    \nonumber \\
    \label{eq:Hproperty}
    H_{mnp}\Gamma^{mnp}\Pi^{(-)}\epsilon
    & =
      \tfrac{1}{8} \left[H_{mnp}\Gamma^{mnp}\epsilon
      +7(\Pi^{(+)}H)_{mnp}\Gamma^{mnp}
      -\Omega_{m}{}^{npq}H_{npq}\Gamma^{m}\right]\epsilon\, .
  \end{align}
\end{subequations}

Finally, we notice the relation

\begin{equation}
  \label{eq:GammamPi-}
  \Gamma^{m}\Pi^{(-)}\epsilon =
  \tfrac{7}{8}\left(\Gamma^{m}-\tfrac{1}{42}\Omega^{m}{}_{npq}\Gamma^{npq}\right)\, .
\end{equation}

\section{Equations of motion at first order in $\alpha'$}
\label{app-EOM}

In this appendix we collect the explicit form of the complete
equations of motion that follow from the action Eq.~(\ref{heterotic})
at first order in $\alpha'$ and ignoring the factor
$g_{s}^{2}/(16\pi G_{N}^{(10)})$.

\begin{eqnarray}
  \label{eq:completedilatoneom}
  \frac{e^{2\phi}}{\sqrt{|g|}}  \frac{\delta S}{\delta \phi}
  & = &
        8\left[\nabla^{2}\phi-(\partial\phi)^{2}\right]
        -2R -\frac{1}{6}H^{2}
        +\frac{\alpha'}{4}\left(F^{2}+R_{(-)}^{2}\right)\, ,
  \\
  & & \nonumber \\
    \label{eq:completevectoreom}
  \frac{1}{\sqrt{|g|}}  \frac{\delta S}{\delta A_{A\, \mu}}
  & = &
        \frac{\alpha'}{2}\left[
        \nabla_{(+)\, \nu}\left(e^{-2\phi}F^{A\, \nu\mu}\right)
        +\frac{1}{2}A^{A}{}_{\nu}
        \nabla_{\rho}\left(e^{-2\phi}H^{\rho\nu\mu}\right)\right]\, ,
  \\
  & & \nonumber \\
    \label{eq:completekalbramondeom}
  \frac{1}{\sqrt{|g|}}  \frac{\delta S}{\delta B_{\mu\nu}}
  & = &
        -\frac{1}{2}\nabla_{\rho}\left(e^{-2\phi}H^{\rho\mu\nu}\right)
        -2\nabla_{\rho}X^{[\rho\mu\nu]}\, ,
  \\
  & & \nonumber \\
    \label{eq:completevielbeineom}
  - \frac{e^{2\phi}}{2\sqrt{|g|}}  \frac{\delta S}{\delta e^{a}{}_{\mu}}
  & = &
        G_{a}{}^{\mu} -2e^{\mu}{}_{a}(\partial\phi)^{2}
        -2\left[\nabla^{\mu}\partial_{a}\phi- e^{\mu}{}_{a}\nabla^{2}\phi \right]
        \nonumber \\
  & & \nonumber \\
  & & 
      +\frac{1}{4}\left[H^{\mu\nu\rho}H_{a\nu\rho}-\frac{1}{6}e^{\mu}{}_{a}H^{2}\right]
      -\frac{\alpha'}{4}\left[F_{A}{}^{\mu\nu}F^{A}{}_{a\nu}-\frac{1}{4}e^{\mu}{}_{a}F^{2}\right]
         \nonumber \\
  & & \nonumber \\
  & & 
      -\frac{\alpha'}{4}\left[
      R_{(-)}{}^{\mu\nu}{}^{b}{}_{c}R_{(-)\, a\nu}{}^{c}{}_{b}
      -\frac{1}{4}e^{\mu}{}_{a}R_{(-)}^{2}\right]
         \nonumber \\
  & & \nonumber \\
  & &
      -\frac{1}{2}e^{2\phi}X^{\rho\nu\mu}H_{\rho\nu a}
      +\frac{1}{2}e^{2\phi}\nabla_{\rho}\left(X^{\rho}{}_{a}{}^{\mu}
      -X^{\mu}{}_{a}{}^{\rho} +X_{a}{}^{\rho\mu}\right)\, ,
\end{eqnarray}

\noindent
where $\nabla_{(+)\, \mu}$ stands for the total (gauge, Lorentz, general
coordinate transformations) covariant derivative with torsionful connection
$\Omega_{(+)}$:

\begin{equation}
\label{eq:eq5}
e^{2\phi}\nabla^{(+)}{}_{\nu}\left(e^{-2\phi}F^{A\, \nu\mu}\right)
=
e^{2\phi}\nabla_{\nu}\left(e^{-2\phi}F^{A\, \nu\mu}\right)
+f_{BC}{}^{A}A^{B}{}_{\nu}F^{C\, \nu\mu}
+\tfrac{1}{2}H_{\nu\rho}{}^{\mu}\wedge{F}^{A\, \nu\rho}
= 
0\, .   
\end{equation}

\noindent
where we have used the shorthand notations

\begin{equation}
  \begin{aligned}
    F^{2}
    & \equiv
    F_{A\, \mu\nu} F^{A\, \mu\nu}\, ,
    \\
    & \\
    R_{(-)}^{2}
    & \equiv
    R_{(-)\, \mu\nu}{}^{a}{}_{b} R_{(-)}{}^{\mu\nu\, b}{}_{a}\, ,
    \\
    & \\
    X^{\nu}{}_{ab}
    & \equiv
    \frac{1}{\sqrt{|g|}}  \frac{\delta S}{\delta \Omega_{(-)\, \nu}{}^{ab}}
    =
    -\frac{\alpha'}{2}\left[ \mathcal{D}_{(+,-)\, \mu}
    \left(e^{-2\phi}R_{(-)}{}^{\mu\nu}{}_{ab}\right)
    +\frac{1}{2}\Omega_{(-)\, \mu\, ab}
    \nabla_{\rho}\left(e^{-2\phi}H^{\rho\mu\nu}\right)
    \right]\, ,
  \end{aligned}
\end{equation}

\noindent
where the last covariant derivative uses the $\Omega_{(+)}$ torsionful
spin connection with respect to the index $\nu$ and $\Omega_{(-)}$
torsionful spin connection with respect to the indices $ab$ and the
Levi-Civita connection with respect to the index $\mu$.

Using the identities

\begin{eqnarray}
  R_{(+)\, abcd}- R_{(-)\, abcd}
  & = &
        2 \nabla_{[a}H_{bcd]}\, ,
  \\
  & & \nonumber \\
  \mathcal{D}_{(\pm)\, [\mu|}R_{(\pm)\, |\nu\rho]}{}^{ab}
  & = &
        0\, ,
\end{eqnarray}

\noindent
where the covariant derivative $\mathcal{D}_{(\pm)\, \mu}$ with
respect to the torsionful spin connections $\Omega_{(\pm)}$ only acts
on the Lorentz indices, and the Bianchi identity of $H$, one can show
that, up to terms of higher order in $\alpha'$ (produced by the
Bianchi identity of $H$)

\begin{equation}
 \mathcal{D}_{(+,-)\, \mu}
 \left(e^{-2\phi}R_{(-)}{}^{\mu c}{}_{ab}\right)
 =
 2e^{-2\phi} \nabla_{(+)\, [a|}\left(R_{(+)\, |b]}{}^{c}
   -2\nabla_{(+)\, |b]}\partial^{c}\phi\right)\, .
\end{equation}

The expression in parenthesis is a combination of the zeroth-order
equations of motion:

\begin{equation}
  R_{(+)\, b}{}^{c}
  -2\nabla_{(+)\, b}\partial^{c}\phi
  =
  R_{b}{}^{c} -2\nabla_{b}\partial^{c}\phi +\frac{1}{4}H^{(0)}{}_{bde}H^{(0)\, cde}
  -\frac{1}{2}e^{2\phi}\nabla_{d}\left(e^{-2\phi}H^{(0)\, d}{}_{b}{}^{c}\right)\, .
\end{equation}

This is the proof of the lemma of Ref.~\cite{Bergshoeff:1989de} mentioned in
Section~\ref{sec-EOM}.

The particular combination of the zeroth-order equations of motion that
appears in the above expression is, in the notation introduced in
Section~\ref{sec-EOM} of the main text,

\begin{equation}
  R_{(+)\, b}{}^{c}
  -2\nabla_{(+)\, b}\partial^{c}\phi
  =
  \frac{e^{2\phi}}{2\sqrt{|g|}}
  \left[\mathcal{E}^{(0)}_{e\, b}{}^{c}+2\mathcal{E}^{(0)}_{B\, b}{}^{c}
    +\tfrac{1}{2}g_{b}{}^{c}\mathcal{E}^{(0)}_{\phi}\right]\, ,
\end{equation}

\noindent
and, therefore,

\begin{equation}
  \label{eq:Xcombination}
    X^{\nu}{}_{ab}
    =
    -\frac{\alpha'}{2\sqrt{|g|}}
    \left\{ e^{-2\phi}
      \nabla^{(0)}_{(+)\, [a|}
      \left[
        e^{2\phi}
  \left(\mathcal{E}^{(0)}_{e\, |b]}{}^{\nu}+2\mathcal{E}^{(0)}_{B\, |b]}{}^{\nu}
    +\tfrac{1}{2}e_{|b]}{}^{\nu}\mathcal{E}^{(0)}_{\phi}
  \right)
        \right]
    -\Omega^{(0)}_{(-)\, \mu\, ab}\mathcal{E}^{(0)}_{B}{}^{\mu\nu}
    \right\}\, ,
\end{equation}

\noindent
where the upper $(0)$ indices indicate zeroth order in $\alpha'$.

\subsection{Noether identities}
\label{app-noether}

In order to understand better the structure of the equations of motion, it is
convenient to study the Noether identities that relate them as a consequence
of the gauge symmetries of the theory.  Associated to the standard gauge
transformations of the Kalb-Ramond 2-form with parameter $\Lambda_{\mu}$

\begin{equation}
\delta B_{\mu\nu} = 2\partial_{[\mu}\Lambda_{\nu]}\, ,  
\end{equation}

\noindent
we find

\begin{equation}
  \label{eq:noetheridentity1}
\partial_{\nu}\frac{\delta S}{\delta B_{\nu\rho}} = 0\, .  
\end{equation}

Associated to the invariance under Yang-Mills gauge transformations with gauge
parameters $\xi^{A}$

\begin{equation}
  \begin{aligned}
    \delta A^{A}{}_{\mu}
    & =
    \mathcal{D}_{\mu}\xi^{A}\, ,
    \\
    & \\
    \delta B_{\mu\nu}
    & =
    -\frac{\alpha'}{2}A_{A\, [\mu}\partial_{\nu]}\xi^{A}\, ,
  \end{aligned}
\end{equation}

\noindent
we find 

\begin{equation}
  \label{eq:noetheridentity2}
  \mathcal{D}_{\mu} \frac{\delta S}{\delta A^{A}{}_{\mu}}
  -\frac{\alpha'}{2} \partial_{\nu}
  \left(A_{A\, \mu} \frac{\delta S}{\delta B_{\mu\nu}} \right)
  =
  0\, .
\end{equation}

If this identity is true, then, this one is also true as well:

\begin{equation}
    \label{eq:noetheridentity3}
  \mathcal{D}_{(-)\, \mu} \frac{\delta S}{\delta \Omega_{(-)\, \mu}{}^{ab}}
  +\frac{\alpha'}{2} \partial_{\nu}
  \left(\Omega_{(-)\, \mu\, ab} \frac{\delta S}{\delta B_{\mu\nu}} \right)
  =
  0\, .
\end{equation}

Associated to the invariance under local Lorentz transformations with
gauge parameters $\sigma^{ab}$

\begin{equation}
  \begin{aligned}
    \delta e^{a}{}_{\mu}
    & =
    \sigma^{a}{}_{b}e^{b}{}_{\mu}\, ,
    \\
    & \\
    \delta \Omega_{(-)\, \mu}{}^{ab}
    & =
    \mathcal{D}_{(-)\, \mu}\sigma^{ab}\, ,
    \\
    & \\
    \delta B_{\mu\nu}
    & =
    -\frac{\alpha'}{2} \Omega_{(-)\, [\mu|}{}^{a}{}_{b}
    \partial_{|\nu]}\sigma^{b}{}_{a}\, ,
  \end{aligned}
\end{equation}

\noindent
we find

\begin{equation}
    \label{eq:noetheridentity4}
  \left.\frac{\delta S}{\delta e^{[a}{}_{\mu}}\right|_{\rm exp}e_{b]\, \mu}
-\mathcal{D}_{(-)\, \mu} \frac{\delta S}{\delta \Omega_{(-)\, \mu}{}^{ab}}
  -\frac{\alpha'}{2} \partial_{\nu}
  \left(\Omega_{(-)\, \mu\, ab} \frac{\delta S}{\delta B_{\mu\nu}} \right)
  =
  0\, ,
\end{equation}

\noindent
which can be simplified with Eq.~(\ref{eq:noetheridentity3}) to

\begin{equation}
    \label{eq:noetheridentity5}
  \left.\frac{\delta S}{\delta e^{[a}{}_{\mu}}\right|_{\rm exp}e_{b]\, \mu}
  =
  0\, .
\end{equation}



\begin{thebibliography}{99}

\bibitem{Tseytlin:1996as}
A.~A.~Tseytlin,
``Extreme dyonic black holes in string theory,''
Mod.\ Phys.\ Lett.\ A {\bf 11} (1996) 689.
\doi{10.1142/S0217732396000709}
[\hepth{9601177}].

\bibitem{Cvetic:1995bj}
M.~Cvetic and A.~A.~Tseytlin,
``Solitonic strings and BPS saturated dyonic black holes,''
Phys.\ Rev.\ D {\bf 53} (1996) 5619.
Erratum: [Phys.\ Rev.\ D {\bf 55} (1997) 3907].
\doi{10.1103/PhysRevD.53.5619}, \doi{10.1103/PhysRevD.55.3907}
[\hepth{9512031}].

\bibitem{Strominger:1996sh}
A.~Strominger and C.~Vafa,
``Microscopic origin of the Bekenstein-Hawking entropy,''
Phys.\ Lett.\ B {\bf 379} (1996) 99.
\doi{10.1016/0370-2693(96)00345-0}
[\hepth{9601029}].

\bibitem{Gibbons:1982fy}
 G.~W.~Gibbons and C.~M.~Hull,
``A Bogomolny Bound for General Relativity and Solitons in N=2 Supergravity,''
Phys.\ Lett.\  {\bf 109B} (1982) 190.
\doi{10.1016/0370-2693(82)90751-1}

\bibitem{Tod:1983pm}
K.~P.~Tod,
``All Metrics Admitting Supercovariantly Constant Spinors,''
Phys.\ Lett.\  {\bf 121B} (1983) 241.
\doi{10.1016/0370-2693(83)90797-9}

\bibitem{Caldarelli:2003pb}
M.~M.~Caldarelli and D.~Klemm,
``All supersymmetric solutions of N=2, D = 4 gauged supergravity,''
JHEP {\bf 0309} (2003) 019.
\doi{10.1088/1126-6708/2003/09/019}
[\hepth{0307022}].

\bibitem{Meessen:2006tu}
 P.~Meessen and T.~Ort\'{\i}n,
``The Supersymmetric configurations of N=2, D=4 supergravity coupled to vector supermultiplets,''
Nucl.\ Phys.\ B {\bf 749} (2006) 291.
\doi{10.1016/j.nuclphysb.2006.05.025}
[\hepth{0603099}].

\bibitem{Huebscher:2006mr}
M.~H\"ubscher, P.~Meessen and T.~Ort\'{\i}n,
``Supersymmetric solutions of N=2 D=4 sugra: The Whole ungauged shebang,''
Nucl.\ Phys.\ B {\bf 759} (2006) 228.
\doi{10.1016/j.nuclphysb.2006.10.004}.
[\hepth{0606281}].

\bibitem{Cacciatori:2008ek}
S.~L.~Cacciatori, D.~Klemm, D.~S.~Mansi and E.~Zorzan,
``All timelike supersymmetric solutions of N=2, D=4 gauged supergravity coupled to abelian vector multiplets,''
JHEP {\bf 0805} (2008) 097.
\doi{10.1088/1126-6708/2008/05/097}
[\arxiv{0804.0009} [hep-th]].

\bibitem{Hubscher:2008yz}
M.~H\"ubscher, P.~Meessen, T.~Ort\'{\i}n and S.~Vaul\`a,
``N=2 Einstein-Yang-Mills's BPS solutions,''
JHEP {\bf 0809} (2008) 099.
\doi{10.1088/1126-6708/2008/09/099}
[\arxiv{0806.1477} [hep-th]].

\bibitem{Klemm:2009uw}
D.~Klemm and E.~Zorzan,
``All null supersymmetric backgrounds of N=2, D=4 gauged supergravity coupled to abelian vector multiplets,''
Class.\ Quant.\ Grav.\  {\bf 26} (2009) 145018.
\doi{10.1088/0264-9381/26/14/145018}
[\arxiv{0902.4186} [hep-th]].

\bibitem{Klemm:2010mc}
D.~Klemm and E.~Zorzan,
``The timelike half-supersymmetric backgrounds of N=2,
  D=4 supergravity with Fayet-Iliopoulos gauging,''
Phys.\ Rev.\ D {\bf 82} (2010) 045012.
\doi{10.1103/PhysRevD.82.045012}
[\arxiv{1003.2974} [hep-th]].

\bibitem{Meessen:2012sr}
P.~Meessen and T.~Ort\'{\i}n,
``Supersymmetric solutions to gauged N=2 d=4 sugra: the full timelike shebang,''
Nucl.\ Phys.\ B {\bf 863} (2012) 65.
\doi{10.1016/j.nuclphysb.2012.05.023}
[\arxiv{1204.0493} [hep-th]].

\bibitem{Gauntlett:2002nw}
J.~P.~Gauntlett, J.~B.~Gutowski, C.~M.~Hull, S.~Pakis and H.~S.~Reall,
``All supersymmetric solutions of minimal supergravity in five- dimensions,''
Class.\ Quant.\ Grav.\  {\bf 20} (2003) 4587.
\doi{10.1088/0264-9381/20/21/005}
[\hepth{0209114}].

\bibitem{Tod:1995jf}
K.~P.~Tod,
``More on supercovariantly constant spinors,''
Class.\ Quant.\ Grav.\  {\bf 12} (1995) 1801.
\doi{10.1088/0264-9381/12/7/020}

\bibitem{Bellorin:2005zc}
J.~Bellor\'{\i}n and T.~Ort\'{\i}n,
``All the supersymmetric configurations of N=4, d=4 supergravity,''
Nucl.\ Phys.\ B {\bf 726} (2005) 171.
\doi{10.1016/j.nuclphysb.2005.07.020}
[\hepth{0506056}].
 
\bibitem{Meessen:2010fh}
P.~Meessen, T.~Ort\'{\i}n and S.~Vaul\`a,
``All the timelike supersymmetric solutions of
all ungauged d=4 supergravities,''
JHEP {\bf 1011} (2010) 072.
\doi{10.1007/JHEP11(2010)072}
[\arxiv{1006.0239} [hep-th]].

\bibitem{Gauntlett:2003fk}
J.~P.~Gauntlett and J.~B.~Gutowski,
``All supersymmetric solutions of minimal 
gauged supergravity in five-dimensions,''
Phys.\ Rev.\ D {\bf 68} (2003) 105009.
 Erratum: [Phys.\ Rev.\ D {\bf 70} (2004) 089901]
\doi{10.1103/PhysRevD.70.089901}, \doi{10.1103/PhysRevD.68.105009}
[\hepth{0304064}].

\bibitem{Gutowski:2004yv}
J.~B.~Gutowski and H.~S.~Reall,
``General supersymmetric AdS(5) black holes,''
JHEP {\bf 0404} (2004) 048.
\doi{10.1088/1126-6708/2004/04/048}
[\hepth{0401129}].

\bibitem{Gauntlett:2004qy}
  J.~P.~Gauntlett and J.~B.~Gutowski,
``General concentric black rings,''
Phys.\ Rev.\ D {\bf 71} (2005) 045002.
\doi{10.1103/PhysRevD.71.045002}
[\hepth{0408122}].

\bibitem{Gutowski:2005id}
J.~B.~Gutowski and W.~Sabra,
``General supersymmetric solutions of five-dimensional supergravity,''
JHEP {\bf 0510} (2005) 039.
\doi{10.1088/1126-6708/2005/10/039}
[\hepth{0505185}].

\bibitem{Bellorin:2006yr}
J.~Bellor\'{\i}n, P.~Meessen and T.~Ort\'{\i}n,
``All the supersymmetric solutions of N=1,d=5 ungauged supergravity,''
JHEP {\bf 0701} (2007) 020.
\doi{10.1088/1126-6708/2007/01/020}
[\hepth{0610196}].

\bibitem{Bellorin:2007yp}
J.~Bellor\'{\i}n and T.~Ort\'{\i}n,
``Characterization of all the supersymmetric solutions 
of gauged N=1, d=5 supergravity,''
JHEP {\bf 0708} (2007) 096.
\doi{10.1088/1126-6708/2007/08/096}
[\arxiv{0705.2567} [hep-th]].

\bibitem{Bellorin:2008we}
J.~Bellor\'{\i}n,
``Supersymmetric solutions of gauged five-dimensional 
supergravity with general matter couplings,''
Class.\ Quant.\ Grav.\  {\bf 26} (2009) 195012.
\doi{10.1088/0264-9381/26/19/195012}
[\arxiv{0810.0527} [hep-th]].

\bibitem{Gutowski:2003rg}
J.~B.~Gutowski, D.~Martelli and H.~S.~Reall,
``All Supersymmetric solutions of minimal supergravity in six- dimensions,''
Class.\ Quant.\ Grav.\  {\bf 20} (2003) 5049.
\doi{10.1088/0264-9381/20/23/008}
[\hepth{0306235}].
  
\bibitem{Cariglia:2004kk}
M.~Cariglia and O.~A.~P.~Mac Conamhna,
``The General form of supersymmetric solutions of 
N=(1,0) U(1) and SU(2) gauged supergravities in six-dimensions,''
Class.\ Quant.\ Grav.\  {\bf 21} (2004) 3171.
\doi{10.1088/0264-9381/21/13/006}
[\hepth{0402055}].

\bibitem{Jong:2006za}
D.~Jong, A.~Kaya and E.~Sezgin,
``6D Dyonic String With Active Hyperscalars,''
JHEP \textbf{11} (2006), 047
\doi{10.1088/1126-6708/2006/11/047}
[\hepth{0608034} [hep-th]].

\bibitem{Akyol:2010iz}
M.~Akyol and G.~Papadopoulos,
``Spinorial geometry and Killing spinor equations of 6-D supergravity,''
Class.\ Quant.\ Grav.\  {\bf 28} (2011) 105001.
\doi{10.1088/0264-9381/28/10/105001}
[\arxiv{1010.2632} [hep-th]].

\bibitem{Lam:2018jln}
H.~Het Lam and S.~Vandoren,
``BPS solutions of six-dimensional (1, 0) supergravity coupled to tensor multiplets,''
JHEP {\bf 1806} (2018) 021
\doi{10.1007/JHEP06(2018)021}
[\arxiv{1804.04681} [hep-th]].

\bibitem{Cano:2019gqm}
P.~A.~Cano and T.~Ort\'{\i}n,
``The structure of all the supersymmetric solutions of ungauged $\mathcal{N} = (1,0),d=6$ supergravity,''
Class.\ Quant.\ Grav.\  {\bf 36} (2019) no.12,  125007
\doi{10.1088/1361-6382/ab1f1e}
[\arxiv{1804.04945} [hep-th]]

\bibitem{Deger:2010rb}
N.~S.~Deger, H.~Samtleben and \"O.~Sario\v{g}lu,
``On The Supersymmetric Solutions of D=3 Half-maximal Supergravities,''
Nucl.\ Phys.\ B {\bf 840} (2010) 29.
\doi{10.1016/j.nuclphysb.2010.06.020}
[\arxiv{1003.3119} [hep-th]].

\bibitem{deBoer:2014iba}
J.~de Boer, D.~R.~Mayerson and M.~Shigemori,
``Classifying Supersymmetric Solutions in 3D Maximal Supergravity,''
Class.\ Quant.\ Grav.\  {\bf 31} (2014) no.23,  235004.
\doi{10.1088/0264-9381/31/23/235004}
[\arxiv{1403.4600} [hep-th]].

\bibitem{Colgain:2015mta}
E.~\'O Colg\'ain,
``All supersymmetric solutions of 3D U(1)$^3$ gauged supergravity,''
JHEP {\bf 1511} (2015) 116.
\doi{10.1007/JHEP11(2015)116}
[\arxiv{1502.04668} [hep-th]].
  
\bibitem{Deger:2015tra}
N.~S.~Deger, G.~Moutsopoulos, H.~Samtleben and \"O.~Sario\v{g}lu,
``All timelike supersymmetric solutions of three-dimensional
half-maximal supergravity,''
JHEP {\bf 1506} (2015) 147.
\doi{10.1007/JHEP06(2015)147}
[\arxiv{1503.09146} [hep-th]].

\bibitem{Gran:2008vx}
U.~Gran, J.~Gutowski and G.~Papadopoulos,
``Geometry of all supersymmetric four-dimensional
N = 1 supergravity backgrounds,''
JHEP {\bf 0806} (2008) 102.
\doi{10.1088/1126-6708/2008/06/102}
[\arxiv{0802.1779} [hep-th]].

\bibitem{Ortin:2008wj}
T.~Ort\'{\i}n,
``The Supersymmetric solutions and extensions of ungauged
matter-coupled N=1, d=4 supergravity,''
JHEP {\bf 0805} (2008) 034.
\doi{10.1088/1126-6708/2008/05/034}
[\arxiv{0802.1799} [hep-th]].

\bibitem{Ortin:2015hya}
T.~Ort\'{\i}n,
``Gravity and Strings'', 2nd edition, 
Cambridge University Press, 2015.

\bibitem{Gran:2018ijr}
U.~Gran, J.~Gutowski and G.~Papadopoulos,
``Classification, geometry and applications of supersymmetric backgrounds,''
Phys.\ Rept.\  {\bf 794} (2019) 1.
\doi{10.1016/j.physrep.2018.11.005}
[\arxiv{1808.07879} [hep-th]].

\bibitem{Bergshoeff:1981um}
E.~Bergshoeff, M.~de Roo, B.~de Wit and P.~van Nieuwenhuizen,
``Ten-Dimensional Maxwell-Einstein Supergravity, Its Currents,
and the Issue of Its Auxiliary Fields,''
Nucl.\ Phys.\ B {\bf 195} (1982) 97.
\doi{10.1016/0550-3213(82)90050-5}

\bibitem{Chapline:1982ww}
G.~F.~Chapline and N.~S.~Manton,
``Unification of Yang-Mills Theory and Supergravity in Ten-Dimensions,''
Phys.\ Lett.\ B {\bf 120} (1983) 105
\doi{10.1016/0370-2693(83)90633-0}

\bibitem{Bergshoeff:1988nn}
E.~Bergshoeff and M.~de Roo,
``Supersymmetric Chern-simons Terms in Ten-dimensions,''
Phys.\ Lett.\ B {\bf 218} (1989) 210.
\doi{10.1016/0370-2693(89)91420-2}
 
\bibitem{Bergshoeff:1989de}
E.~A.~Bergshoeff and M.~de Roo,
``The Quartic Effective Action of the Heterotic String and Supersymmetry,''
Nucl.\ Phys.\ B {\bf 328} (1989) 439.
\doi{10.1016/0550-3213(89)90336-2}

\bibitem{Gran:2005wf}
U.~Gran, P.~Lohrmann and G.~Papadopoulos,
``The Spinorial geometry of supersymmetric heterotic string backgrounds,''
JHEP {\bf 0602} (2006) 063.
\doi{10.1088/1126-6708/2006/02/063}
[\hepth{0510176}].

\bibitem{Gran:2007fu}
U.~Gran, G.~Papadopoulos, D.~Roest and P.~Sloane,
``Geometry of all supersymmetric type I backgrounds,''
JHEP {\bf 0708} (2007) 074.
\doi{10.1088/1126-6708/2007/08/074}
[\hepth{0703143} [hep-th]].
  
\bibitem{Gran:2007kh}
U.~Gran, G.~Papadopoulos and D.~Roest,
``Supersymmetric heterotic string backgrounds,''
Phys.\ Lett.\ B {\bf 656} (2007) 119.
\doi{10.1016/j.physletb.2007.09.024}
[\arxiv{0706.4407} [hep-th]].

\bibitem{Gillard:2004xq}
J.~Gillard, U.~Gran and G.~Papadopoulos,
``The Spinorial geometry of supersymmetric backgrounds,''
Class.\ Quant.\ Grav.\  {\bf 22} (2005) 1033.
\doi{10.1088/0264-9381/22/6/009}
[\hepth{0410155}].

\bibitem{Metsaev:1987zx}
  R.~R.~Metsaev and A.~A.~Tseytlin,
  ``Order alpha-prime
  (Two Loop) Equivalence of the String Equations of Motion and the Sigma Model
  Weyl Invariance Conditions: Dependence on the Dilaton and the Antisymmetric
  Tensor,''
Nucl.\ Phys.\ B {\bf 293} (1987) 385.
\doi{10.1016/0550-3213(87)90077-0}

\bibitem{Kallosh:1993wx}
R.~Kallosh and T.~Ort\'{\i}n,
``Killing spinor identities,''
\hepth{9306085}.

\bibitem{Bellorin:2005hy}
J.~Bellor\'{\i}n and T.~Ort\'{\i}n,
``A Note on simple applications of the Killing Spinor Identities,''
Phys.\ Lett.\ B {\bf 616} (2005) 118.
\doi{10.1016/j.physletb.2005.04.026}
[\hepth{0501246}].


\bibitem{Cortes:2019xmk}
V.~Cort\'es, C.~Lazaroiu and C.~S.~Shahbazi,
``Spinors of real type as polyforms and the generalized Killing equation,''
\arxiv{1911.08658} [math.DG].

\bibitem{Bergshoeff:1990ax}
E.~A.~Bergshoeff and M.~de Roo,
``The string effective action in the dual formulation of D = 10 supergravity,''
Phys.\ Lett.\ B {\bf 247} (1990) 530.
\doi{10.1016/0370-2693(90)91896-J}

\bibitem{Bergshoeff:1990hh}
E.~A.~Bergshoeff and M.~de Roo,
``Duality transformations of string effective actions,''
Phys.\ Lett.\ B {\bf 249} (1990) 27.
\doi{10.1016/0370-2693(90)90522-8}

\bibitem{Gross:1986iv}
D.~J.~Gross and E.~Witten,
``Superstring Modifications of Einstein's Equations,''
Nucl.\ Phys.\ B {\bf 277} (1986) 1.
\doi{10.1016/0550-3213(86)90429-3}

\bibitem{Grisaru:1986dk}
M.~T.~Grisaru, A.~E.~M.~van de Ven and D.~Zanon,
``Two-Dimensional Supersymmetric Sigma Models on Ricci Flat Kahler
Manifolds Are Not Finite,''
Nucl.\ Phys.\ B {\bf 277} (1986) 388.
\doi{10.1016/0550-3213(86)90448-7}

\bibitem{Grisaru:1986kw}
M.~T.~Grisaru, A.~E.~M.~van de Ven and D.~Zanon,
``Four Loop Divergences for the N=1 Supersymmetric Nonlinear Sigma Model in
 Two-Dimensions,''
Nucl.\ Phys.\ B {\bf 277} (1986) 409.
\doi{10.1016/0550-3213(86)90449-9}

\bibitem{Fubini:1985jm}
S.~Fubini and H.~Nicolai,
``The Octonionic Instanton,''
Phys.\ Lett.\  {\bf 155B} (1985) 369.
\doi{10.1016/0370-2693(85)91589-8}

\bibitem{Harvey:1990eg}
J.~A.~Harvey and A.~Strominger,
``Octonionic superstring solitons,''
Phys.\ Rev.\ Lett.\  {\bf 66} (1991) 549.
\doi{10.1103/PhysRevLett.66.549}

\bibitem{Strominger:1990et}
A.~Strominger,
``Heterotic solitons,''
Nucl.\ Phys.\ B {\bf 343} (1990) 167.
Erratum: [Nucl.\ Phys.\ B {\bf 353} (1991) 565].
\doi{10.1016/0550-3213(91)90349-3}, \doi{10.1016/0550-3213(90)90599-9}

\bibitem{Belavin:1975fg}
A.~A.~Belavin, A.~M.~Polyakov, A.~S.~Schwartz and Y.~S.~Tyupkin,
``Pseudoparticle Solutions of the Yang-Mills Equations,''
Phys.\ Lett.\ B {\bf 59} (1975) 85.
\doi{10.1016/0370-2693(75)90163-X}.

\bibitem{Gunaydin:1995ku}
M.~Gunaydin and H.~Nicolai,
``Seven-dimensional octonionic Yang-Mills instanton and
its extension to an heterotic string soliton,'' 
Phys.\ Lett.\ B {\bf 351} (1995) 169
Addendum: [Phys.\ Lett.\ B {\bf 376} (1996) 329]
\doi{10.1016/0370-2693(95)00375-U}
[\hepth{9502009}].

\bibitem{Bergshoeff:1992cw}
E.~A.~Bergshoeff, R.~Kallosh and T.~Ort\'{\i}n,
``Supersymmetric string waves,''
Phys.\ Rev.\ D {\bf 47} (1993) 5444.
\doi{10.1103/PhysRevD.47.5444}
[\hepth{9212030}].

\bibitem{Deser:1976eh}
S.~Deser and B.~Zumino,
``Consistent Supergravity,''
Phys.\ Lett.\  {\bf 62B} (1976) 335.
\doi{10.1016/0370-2693(76)90089-7}

\bibitem{Meessen:2007ef}
P.~Meessen,
``All-order consistency of 5d sugra vacua,''
Phys.\ Rev.\ D {\bf 76} (2007) 046006.
Erratum: [Phys.\ Rev.\ D {\bf 85} (2012) 129902].
\doi{10.1103/PhysRevD.85.129902}, \doi{10.1103/PhysRevD.76.046006}
[\arxiv{0705.1966} [hep-th]].

\bibitem{kn:FO}
  A.~Fontanella and T.~Ort\'{\i}n,
  work in progress.

\bibitem{kn:CSS}
  C.S.~Shahbazi,
  ``Spin(7) structures from chiral spinors on
  ten-dimensional Lorentzian spin manifolds,''
  to appear.
  
\bibitem{Duff:1990wv}
M.~J.~Duff and J.~X.~Lu,
``Elementary five-brane solutions of D = 10 supergravity,''
Nucl.\ Phys.\ B {\bf 354} (1991) 141.
\doi{10.1016/0550-3213(91)90180-6}

\bibitem{Callan:1991dj}
C.~G.~Callan, Jr., J.~A.~Harvey and A.~Strominger,
``World sheet approach to heterotic instantons and solitons,''
Nucl.\ Phys.\ B {\bf 359} (1991) 611.
\doi{10.1016/0550-3213(91)90074-8}

\bibitem{kn:Joyce}
  D.D.~Joyce,
  ``Compact Manifolds with Special Holonomy,''
  Oxford Mathematical Monographs,
  Oxford University Press,
  Oxford, U.K. (2000)

\bibitem{deWit:1983gs}
B.~de Wit and H.~Nicolai,
``The Parallelizing S(7) Torsion in Gauged $N=8$ Supergravity,''
Nucl.\ Phys.\ B {\bf 231} (1984) 506.
\doi{10.1016/0550-3213(84)90517-0}




















\end{thebibliography}
\end{document}